\documentclass[fleqn,usenatbib]{mnras}

\usepackage{newtxtext,newtxmath}

\usepackage[T1]{fontenc}
\usepackage{ae,aecompl}

\DeclareRobustCommand{\VAN}[3]{#2}
\let\VANthebibliography\thebibliography
\def\thebibliography{\DeclareRobustCommand{\VAN}[3]{##3}\VANthebibliography}

\usepackage{graphicx}	
\usepackage{amsmath}	
\usepackage{amssymb}	

\newcommand{\mum}{\,$\mu$m\,}	
\newcommand{\kms}{\,km\,s$^{-1}$\,} 
\newcommand{\tauboo}{$\tau$ Bo\"otis\,}

\title[Atmospheric water detection for \tauboo b]{Water observed in the atmosphere of \tauboo Ab with CARMENES/CAHA}

\author[R. K. Webb et al.]{
Rebecca K. Webb,$^{1,2}$\thanks{E-mail: r.k.webb@warwick.ac.uk}
Siddharth Gandhi,$^{3,1,2}$
Matteo Brogi,$^{1,2,4}$
Jayne L. Birkby,$^{5}$
\newauthor
Ernst de Mooij,$^{6}$
Ignas Snellen$^{3}$
and Yapeng Zhang$^{3}$
\\
$^{1}$Department of Physics, University of Warwick, Gibbet Hill Road, Coventry, CV4 7AL, UK\\
$^{2}$Centre for Exoplanets and Habitability, University of Warwick, Gibbet Hill Road, Coventry, CV4 7AL, UK\\
$^{3}$Leiden Observatory, Leiden University, Postbus 9513 2300 RA, Leiden, The Netherlands\\
$^{4}$INAF-Osservatorio Astrofisico di Torino, Via Osservatorio 20, I-10025, Pino Torinese, Italy\\
$^{5}$Astrophysics, University of Oxford, Denys Wilkinson Building, Keble Road, Oxford OX1 3RH, UK\\
$^{6}$Astrophysics Research Centre, Queen's University Belfast, Belfast BT7 1NN, UK\\
}

\date{Accepted XXX. Received YYY; in original form ZZZ}

\pubyear{2015}

\begin{document}
\label{firstpage}
\pagerange{\pageref{firstpage}--\pageref{lastpage}}
\maketitle

\begin{abstract}
Characterising the atmospheres of hot Jupiters is important in understanding the formation and migration of these exotic planets. However, there are still many open questions about the chemical and physical properties of these atmospheres. Here, we confirm the detection of water vapour in thermal emission from the non-transiting hot Jupiter \tauboo Ab with the high resolution NIR CARMENES spectrograph. Combining over 17 h of observations (560 spectra) and using a Bayesian cross-correlation to log-likelihood approach, 
we measure a systemic velocity of $V_{\mathrm{sys}} = -11.51^{+0.59}_{-0.60}$\,\kms and a radial velocity semi-amplitude of $K_{\mathrm{P}} = 106.21^{+1.76}_{-1.71}$\,\kms for the planet, which results in an absolute mass of $M_{\mathrm{P}} = 6.24^{+0.17}_{-0.18}\,\mathrm{M_{J}}$ and an orbital inclination of $41.6^{+1.0}_{-0.9}$ degrees. 
Our retrieved $V_{\mathrm{sys}}$ shows a significant shift (+5 \kms) from the literature value, which could be caused by an inaccurate time of periastron.
Within the explored model grid, we measure a preference for solar water abundance (VMR = $10^{-3}$) and no evidence for additional minor species in the atmosphere.
Given the extensive orbital coverage of the data, we searched for a phase dependency in the water signal but found no strong evidence of variation with orbital phase.
This detection is at odds with recent observations from SPIRou/CFHT and their tight upper limit on water abundance. We recommend further observations of the atmosphere \tauboo Ab to try and resolve these discrepancies.
\end{abstract}




\section{Introduction} \label{sec: Intro}

Having been discovered by \cite{Butler1997}, \tauboo b was one of the first main-sequence exoplanets to be observed with precise ($\sim$\,m\,s$^{-1}$) Doppler shift measurements of stellar absorption lines using high resolution echelle spectrographs. These first generation of exoplanet discoveries became known as hot Jupiters due to their large (Jupiter-like) masses and close-in orbits that are a fraction of the orbit of Mercury in the Solar system, which results in high day-side temperatures for these planets ($T_{\mathrm{P}}$\,$\geq$\,$1000$\,K). Due to these extreme conditions but favourable signals, these exoplanets are also the most ideal candidates for atmospheric characterisation despite accounting for $\lesssim1$\,per cent of sun-like stars \citep[e.g.][]{Wright2012,Petigura2018,Boley2021}. \tauboo b orbits a hot, bright F7 \citep{Gray2001} star ($V = 4.49$, $H = 3.55$) \citep{vanBelle2009,Cutri2003} that is part of a binary system with an M3 dwarf companion \citep{Joy1974}. Follow up photometric observations found that the planet was not to be transiting \cite{Baliunas1997}. 

The close proximity of \tauboo b to one of the brightest planetary host stars has made this system one of the most observed non-transiting planets over the last couple of decades. Searches for the reflected light from \tauboo b using high resolution instruments commenced soon after its discovery  \citep{Charbonneau1999,Collier-Cameron1999} to directly detect the atmospheric composition, geometric albedo, absolute mass and the orbital inclination. Successive observations \citep{Leigh2003,Rodler2010} have also been unsuccessful in detecting the reflected light from the atmosphere of \tauboo b. \cite{Hoeijmakers2018} produced a meta-analysis of all the data determining an upper-limit on the planet-to-star contrast and the optical albedo of $1.5\times10^{-5}$ and 0.12, respectively. Recent radial velocity measurements of \tauboo A have constrained the orbital parameters of the M-dwarf companion and the planet \citep{Justesen2019}, the former of which is expected to reach periastron by $\sim2026$. By this time, the M-dwarf is likely to be within $1''$ of the A star which could cause significant contamination for spectroscopic measurements of the host star and the planet. 

The first direct detections from an atmosphere of a non-transiting hot Jupiter was observed from \tauboo b in the near-infrared (NIR) \citep{Brogi2012,Rodler2012} with the high resolution CRIRES instrument \citep{CRIRES} at the VLT. At $\sim2.3$\mum, they were able to accurately constrain the radial velocity of the planet by tracing the thermal emission of CO around superior-conjunction where the hot day-side comes into view. Later, \cite{Lockwood2014} reported the first detection of water vapour in the atmosphere of \tauboo b in the \textit{L}-band with a $6\,\sigma$ detection from NIRSPEC/Keck \citep{NIRSPEC}. Over the past decade, the use of high resolution spectroscopy has been optimal for measuring the chemical composition \citep[e.g.][]{Giacobbe2021}, wind speeds \citep[e.g.][]{Louden2015} and temperature-pressure ($T-p$) profiles \citep[e.g.][]{Brogi2014}. More recently, developments have been made on a Bayesian framework to enable a full atmospheric retrieval from these high resolution spectra \citep{Brogi2017,Brogi2019,Gibson2020,Nugroho2020}. This has allowed for the retrieval of precise absolute chemical abundances \citep{Pelletier2021} and the C/O ratio \citep{Line2021}. 
The result from \citet{Pelletier2021} is particularly interesting in this context, as only carbon monoxide was measured in the atmosphere of \tauboo\,Ab. In spite of the precise abundance derived for this species ($\log$(CO)$=-2.46^{+0.25}_{-0.29}$), they reported a non-detection of water vapour and placed a tight upper limit of VMR $\log$(H$_2$O)$\le-5.66\ (3\sigma)$. Given the previous detection of water by \citet{Lockwood2014} and the inconclusive evidence from \citet{Brogi2012}, \tauboo\ is an interesting planet to revisit with independent data, which is what we present in this paper.

The bulk of NIR observations of hot Jupiters thus far have been to measure the thermal emission brightness variation of the planet as a function of line of sight (i.e a phase curve). This gives a 3-dimensional picture of the temperature and cloud coverage across the longitude of the planet \citep[e.g.][]{Knutson2007,Kreidberg2018}. These phase variations leaves imprints in their chemical and $T-p$ structure as a function of the orbital phase \citep[e.g.][]{Stevenson2017}. These phase variations are expected to become more prominent for the hotter class of hot Jupiters known as the ultra-hot Jupiters ($>2000\,$K) \citep{Parmentier2018} because of the extreme temperature gradient between the hot day-side and cooler night-side. This variation in the temperature structure causes thermal dissociation to occur which plays an important role in the heat recirculation of the hottest planets \citep{Bell2018}. \cite{Ehrenreich2020} and \cite{Kesseli2021} measured the phase resolved variation in the absorption of neutral iron in the transmission spectrum of the ultra-hot Jupiter WASP-76b, attributed to the rain-out of iron on the night-side, with high resolution spectroscopy. However, at high resolution, phase variations in the thermal emission spectrum have remained elusive. 

In this analysis, we confirm the detection of water vapour in the thermal emission with day-side observations of the hot Jupiter \tauboo b with high resolution spectroscopy. \tauboo b is estimated to have day-side temperature limits between $\sim1980$ 
and 1670\,K 
assuming zero and perfect heat redistribution to the night-side, respectively \citep{Cowan2011}. We also split the spectra into pre- and post-superior conjunction to measure any variations in the water signal as a function of phase. We outline the observations in section~\ref{sec: Observations}, the data reduction and analysis in section~\ref{sec: data analysis}, the results in sections~\ref{sec: tell removal} and \ref{sec: results}. Finally, we discuss the results in section~\ref{sec: discussion} with a summary of our conclusions in section~\ref{sec: conclusions}.  

\section{Observations} \label{sec: Observations}

\begin{figure}
    \centering
    \includegraphics[width=\columnwidth]{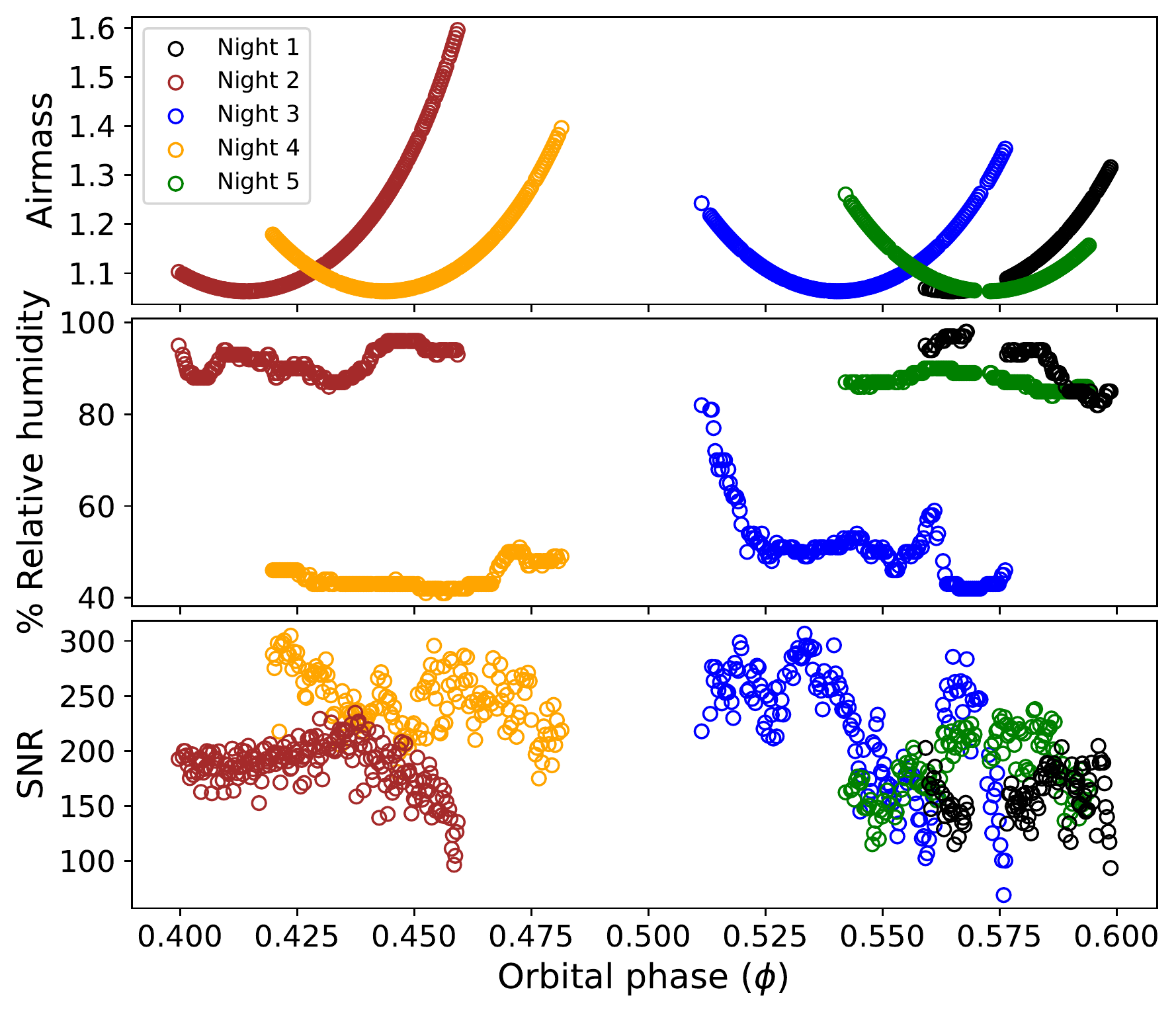}
    \caption{The airmass (top panel), relative humidity (middle panel) and the SNR of order 51 ($\lambda = 1.19-1.21$\mum) as a function orbital phase for the observations of \tauboo b of spectra taken with CARMENES.}
    \label{fig: observations}
\end{figure}

\begin{table*}
    \centering
    \caption{Observations of \tauboo with CAMRNES. The SNR was calculated as the average SNR of order 51 ($\lambda = 1.19-1.21$\mum) over the night. The average $V_{\mathrm{P}}$ is calculated from equation~\ref{eq: planet velocity} assuming a planet $K_{\mathrm{P}}$ of 110\kms and a $V_{\mathrm{sys}}$ of -16.9\kms averaged over the entire night.}
    \begin{tabular}{lcccccc}
    \hline
    \textbf{Night} & \textbf{Date} & \textbf{Number of spectra} & \textbf{Exposure time (s)} & \textbf{Phase range} & \textbf{SNR} & \textbf{Average $V_{\mathrm{P}}$ (\kms)}\\
    \hline
    1 & 2018-March-26 & 110 & 40 & 0.559-0.599 & 148 & -76.6 \\
    2 & 2018-May-11 & 261 & 40 & 0.399-0.459 & 165 & 44.3 \\
    3 & 2019-March-12 & 161 & 70 & 0.511-0.576 & 222 & -58.4 \\
    4 & 2019-March-15 & 165 & 65 & 0.420-0.481 & 246 & 5.74 \\
    5 & 2019-April-11 & 133 & 66 & 0.542-0.594 & 186 & -61.4 \\
    \hline
    \end{tabular}
    \label{tab: info on obs}
\end{table*}

The day-side emission from the $\tau$ Bo\"otis Ab system was observed over five nights (see also \cite{Zhang2020}) with the high resolution spectrograph CARMENES mounted on the 3.5\,-\,m telescope at the Calar Alto Observatory \citep{CARMENES}. We used two nights of spectra taken from the Calar Alto archive (PI: J.A.Caballero and F.J.Alonso-Floriano) which were taken on 2018 March 26 and 2018 May 11 (nights 1 and 2, hereafter). We also observed (PI: M.Brogi) \tauboo b for a further three nights on 2019 March 12, 2019 March 15 and 2019 April 11 (nights 3, 4 and 5, hereafter). CARMENES consists of  separate optical and NIR spectrographs with spectral resolutions of $R=94,600$ and 80,400, respectively. In this analysis, we utilise the NIR spectra to probe the thermal emission from the planet with a wavelength coverage of $\lambda = 0.96 - 1.71$\mum which spans the \textit{Y}, \textit{J} and \textit{H}-bands with a sampling precision of $\sim3.7\,\mathrm{km\,s^{-1}\,pixel^{-1}}$. The NIR spectra are dispersed into 28 echelle orders, orders 63\,-\,36, which are imaged on two 2k$\times$2k Hawaii-2RG detectors.

A time series of spectra was taken pre- ($\phi$\,$<$\,0.5, nights 2 and 4) and post superior-conjunction ($\phi$\,$>$\,0.5, nights 1, 3 and 5). In table~\ref{tab: info on obs}, we show the observations of \tauboo taken with CARMENES with their respective average exposure times, number of spectra obtained and observed phase range. For nights 1 and 2, we removed the final 10 and 52 spectra, respectively, due to a rapidly decreasing SNR in the spectra likely due to increased cloud coverage. For each exposure, one fibre was held on the target and a second fibre was placed on the sky for accurate thermal background subtraction. Each spectrum was subsequently calibrated and reduced using the \textsc{caracal} v2.10 \citep{CARACAL} pipeline which performs a dark/bias correction, order tracing, a flat-relative optimal extraction and an accurate wavelength calibration from a U-Ne lamp. 
We use this solution - which is known to be stable at the m\,s$^{-1}$ level, for our analysis. 
Thus, we do not rebin the spectra onto a grid at constant resolving power, nor do we apply any correction for shifts at the sub-pixel level.

From Fig.~\ref{fig: observations}, night 4 had the best observing conditions out of the five with the other nights suffering from large variability in atmospheric conditions and high humidity levels. This is reflected in the most stable signal-to-noise ratio (SNR) for all wavelengths (or orders) as shown in the bottom panel of Fig.~\ref{fig: observations}. 
We note that while the relative humidity at the level of the telescope is not an exact proxy for the precipitable water vapour over the entire atmospheric column density, it is at least an indicator of the overall quality of the night.

\section{Telluric removal} \label{sec: tell removal}

\begin{figure*}
    \centering
    \includegraphics[width=\textwidth]{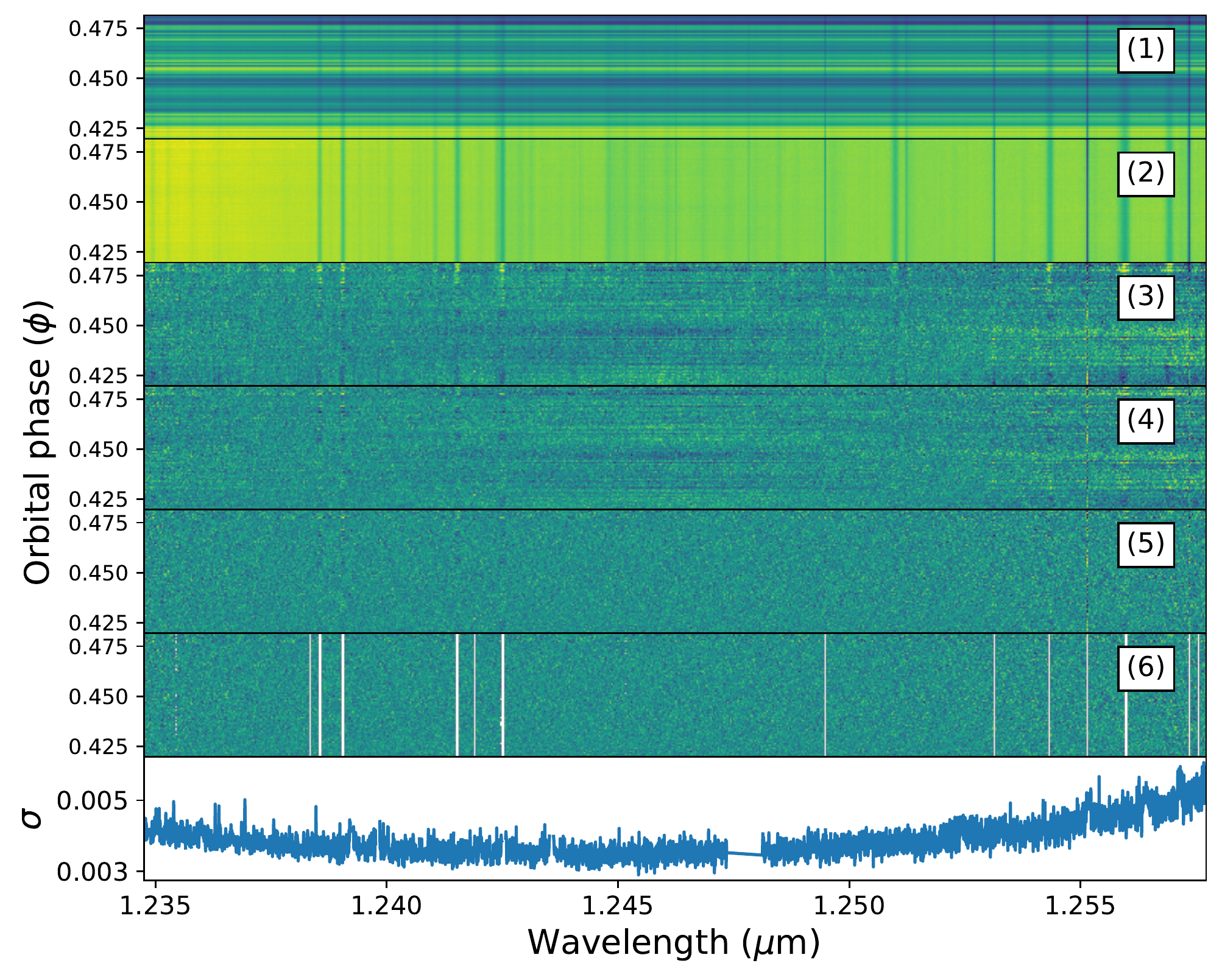}
    \caption{Example of the telluric removal sequence on order 49 of the NIR arm of CARMENES for night 4. The panel numbers indicate the telluric removal steps as described in the text. The bottom panel shows the final standard deviation in time for the spectra. The contrast is increased in panels 3-6 for better visualisation of the noise structure in the spectral sequence.}
    \label{fig: tellrem steps}
\end{figure*}

Firstly, we need to remove the contaminating telluric and host stellar absorption that dominate the extracted spectra. For each order, the time resolved spectra are set-up in 2-D matrices with dimensions $\mathrm{n_{f}}$\,$\times$\,$\mathrm{n_{x}}$, where $\mathrm{n_{f}}$ is the frame number and $\mathrm{n_{x}}$ is the number of pixels or wavelength channel, for CARMENES detectors this is 4080. Before passing the data through the telluric pipeline, we removed orders 41-46 ($\lambda = 1.32-1.50$\mum), 52-55 ($\lambda = 1.10-1.19$\mum) as these were saturated with telluric lines.   

We remove the variation in the observed flux as a function of time. As the telluric features are stationary over a nights observations, we can effectively treat each wavelength channel as a light curve which can then be detrended from telluric depth variations. Since these spectra were taken at orbital phases close to superior conjunction, we are observing the maximum rate of change in the radial velocity shift of the planet ($\frac{\mathrm{d}V_{P}}{\mathrm{d}t}$) thereby minimising the losses of the planetary signature which will be Doppler shifted across several wavelength channels over the time series. 
Among the various techniques to remove telluric lines, in this analysis we primarily follow the algorithm used in \cite{Brogi2019} and \cite{Webb2020}. Fig.~\ref{fig: tellrem steps} shows an example of the 2-D time resolved spectra as a function of wavelength for the CARMENES echelle order 49 for night 4. Each panel shows each step in the reduction of the spectra, of which we outline below:

\begin{itemize}
    \item[(1):] Extracted spectra from the \textsc{caracal} calibration pipeline (see Section~\ref{sec: Observations}) stacked in time.
    \item[(2):] The variations in the light throughput are removed by normalising each spectrum (each row in Fig.~\ref{fig: tellrem steps}) by the median of the spectrum.
    \item[(3):] The time-averaged spectrum is fit with a second order polynomial to each observed spectrum. This fit is thus divided through each of the spectra to remove the telluric variability as a function of wavelength.   
    \item[(4):] The telluric lines are further detrended in time for each wavelength channel (each column in Fig.~\ref{fig: tellrem steps}). This is achieved by fitting the spectra in wavelength as a second order polynomial with time (we use the orbital phase as our time stamp here). This function is then divided through for each wavelength channel to remove the telluric variations as a function of time.
    \item[(5):] At this stage, we pass each order through a Gaussian high-pass filter (bin-size of 80 pixels) to remove the wavelength dependent continuum fluctuations which are still present in the data. 
    \item[(6):] The strongest telluric residuals that remained were masked from the analysis. To determine these highly deviant channels, we calculated the ratio between the standard deviation and the relative errors output from the \textsc{caracal} pipeline and thus calculated a median value. A calculation of the difference between this ratio and the median values was then used to determine the channels that rose above the average noise level of the data. Highly deviant pixels are also removed through a 5\,$\sigma$ clipping.   
\end{itemize}

As described above, in step (6) we had to mask telluric residuals that were not corrected for in the telluric removal algorithm and were seen as highly deviant channels in the time-series spectra. This is necessary so as to prevent strong time-correlated telluric noise appearing in the final cross-correlation analysis (see Fig. 4 in \cite{Brogi2018}). The bottom panel in Fig. \ref{fig: tellrem steps} shows the final standard deviation in time as a function of wavelength, the lack of highly deviant wavelength channels indicates that the strongest residuals that were left in the data have now been removed. Over all orders, this resulted in a total of $\sim$\,9, 7, 7, 6 and 7 per cent of the pixels being masked for nights 1-5, respectively. In the case of night 2, we found particularly strong telluric residuals that still remained in the spectra despite this additional masking when correlated with a pure water model (see Fig.~\ref{fig: n5 logl ph bin}. As a result, we remove night 2 from the remainder of the analysis in order to avoid any biases in our results. This is further explained in section~\ref{subsec: single species analysis}. We do however repeat the analysis with the inclusion of night 2 in appendix~\ref{sec: appendix mcmc ind nights} and \ref{app: CC to logl all nights}. 

\section{Data analysis} \label{sec: data analysis}

\subsection{Cross-correlation to log likelihood mapping} \label{subsec: CC and logL mapping}

\begin{table*}
\centering
\caption{Stellar and planetary parameters and their values that was used in the analysis.}
\begin{tabular}{lccc}
\hline
\textbf{\tauboo A} & \textbf{Symbol (units)} & \textbf{Value} & \textbf{Reference}\\
\hline
Spectral type & & F7 & \cite{Gray2001}\\
$H$-band infrared brightness & $H$ (mag) & 3.55 & \cite{Cutri2003}\\ 
Effective temperature & $T_{\mathrm{eff}}$ (K) & 6399\,$\pm$\,45 & \cite{Borsa2015}\\
Mass & $M_{\star}$ ($\mathrm{M}_{\odot}$) & $1.35\pm0.03$ & \cite{Takeda2007}\\
Radius & $R_{\star}$ ($\mathrm{R}_{\odot}$) & 1.42\,$\pm$\,0.08 & \cite{Borsa2015}\\
Luminosity & $L_{\star}$ ($L_{\odot}$) & $3.06\pm0.16$ & \cite{Borsa2015}\\
Distance & $d$ (pc) & $15.66\pm0.08$ & \cite{GaiaDR2}\\
Radial velocity semi-amplitude & $K_{\star}$ ($\mathrm{m\,s^{-1}}$) & $468.42\,\pm2.09$ & \cite{Justesen2019}\\
Systemic velocity & $V_{\mathrm{sys}}$ (\kms) & -16.9\,$\pm$\,0.3 & \cite{GaiaDR2}\\
\hline
\textbf{\tauboo Ab} \\
\hline
Orbital period & $P$ (days) & 3.31245\,$\pm$\,3\,$\times$\,$10^{-6}$ & \cite{Justesen2019}\\
Radius & $R_{\mathrm{P}}$ ($\mathrm{R_{J}}$) & 1.2 & (\textit{estimated})\\
Phase zero-point (in the rest frame of \tauboo A) & $T_{0}$ (MJD) & 56401.879\,$\pm$\,0.004 & \cite{Justesen2019}\\
Orbital separation & $a$ (au) & $0.04869^{+0.00039}_{-0.00040}$ & \cite{Rosenthal2021}\\
Eccentricity & $e$ & $0.0074^{+0.0059}_{-0.0048}$ & \cite{Rosenthal2021}\\
\hline
\end{tabular}
\label{table: tauboo parms}
\end{table*} 

As seen in the residual spectral matrix in panel (5) in Fig.~\ref{fig: tellrem steps}, the planetary signal is still hidden within the noise of the data, i.e the SNR\,$\ll$\,1 per line. To extract this signal, we cross-correlate the observed spectra with model spectra of opacity sources that may be present in the atmosphere. This amplifies the planet SNR by $\sim\sqrt{N_\mathrm{lines}}$, where $N_\mathrm{lines}$ is the number of strong spectral lines.

We Doppler shift the model spectra into the planet rest frame via spline interpolation, and by the radial velocity of the planet in the observer's frame,
\begin{equation}
    V_{\mathrm{P}}(t) = K_{\mathrm{P}}\sin{\{2\pi[\phi(t)+0.5]\}} - V_{\mathrm{bary}}(t) + V\mathrm{_{sys}} , \label{eq: planet velocity}
\end{equation}
where $K_{\mathrm{P}}$ is the radial velocity semi-amplitude of the planet, $V\mathrm{_{bary}}$ and $V\mathrm{_{sys}}$ are the barycentric-earth radial velocity correction and the systemic velocity, respectively. 
The minus sign in $V\mathrm{_{bary}}$ corrects for the fact that barycentric velocities are the observer's velocity computed in the barycentre of the solar system, and not vice-versa as needed here.
The orbital phases are defined as,
\begin{equation}
    \phi(t) = \frac{t - T_0}{P} , \label{eq: orbital phase}
\end{equation}
where $t$ is the time of observations in HJD, $T_{0}$ and $P$ is the phase zero-point in HJD and the orbital period in days, respectively. We note that the $T_{0}$ stated in \cite{Justesen2019} is in MJD we needed to convert $T_{0}$ into HJD using the \texttt{astropy.time} module \citep{Astropy2013, Astropy2018}. The $T_{0}$ was also determined in the rest-frame of the host star, therefore, the orbital phases calculated with this solution will be in the rest-frame of the star. Thus, we need to correct the orbital phases by $0.5$ to be in the rest-frame of \tauboo b, as indicated in equation~\ref{eq: planet velocity}. Before cross-correlation, we must also scale the model spectra to the stellar flux of \tauboo,
\begin{equation} \label{equation: scaled flux}
    F_{\mathrm{scaled}}(\lambda) = \frac{aF_{\mathrm{P}}}{\pi B(\lambda, T\mathrm{_{eff}})}\left(\frac{R_{\mathrm{P}}}{R_{\star}}\right)^{2},
\end{equation}
where $B(\lambda,T_{\mathrm{eff}})$ is the blackbody stellar flux calculated from the measured effective temperature ($T_{\mathrm{eff}}$), $F_{\mathrm{P}}$ is the modelled emergent flux from the planet in W\,m$^{-2}$\,m$^{-1}$, $R_{\mathrm{P}}$ and $R_{\star}$ are the estimated planetary and measured stellar radii, respectively. The scaling factor $a$ in equation~\ref{equation: scaled flux} is assumed to be unity in sections~\ref{subsec: single species analysis} and \ref{subsec: combined species}, however, it is allowed to vary in the MCMC analysis as described in section~\ref{subsec: mcmc}. We assume a simple blackbody for the stellar flux as we do not expect any significant molecular absorption in the stellar atmosphere of \tauboo, which has a temperature of $\sim6400\,$K. The parameter values that were used in the analysis are summarised in table \ref{table: tauboo parms}. 
We adopt a radius of $1.2\,R_{\mathrm{J}}$, which is approximately the radius adopted in \citet{Pelletier2021} ($1.15\,R_{\mathrm{J}}$) corrected by their retrieved scaling factor of $\sim1.04$.
Since \tauboo b is a non-transiting planet, the $K_{\mathrm{P}}$ is uncertain, therefore, we must test a range of velocities of the planet around the expected value of $K_{\mathrm{P}} = 110$\,\kms from the literature \citep{Brogi2012,Lockwood2014}. We tested a range of velocities at $0\,\geq\,K_{\mathrm{P}}\,\geq\,200$ \kms and $-60\,\geq\,V_{\mathrm{sys}}\,\geq\,60$ \kms in steps of 2.0 \kms which is within the velocity resolution of CARMENES ($\sim3.7$\kms in the NIR). The planet radius is also unknown for non-transiting systems; however, we can absorb any uncertainty in the radius into a scaling parameter $\log(a)$, which is then retrieved with the other atmospheric parameters (see section~\ref{subsec: mcmc}).

We now add an additional step into the analysis whereby we include a cross-correlation to log likelihood (CC-$\log(L)$) mapping in order extract statistically robust atmospheric and orbital parameters at high resolution. Since we have time-resolved spectra on a timescale where any planet signal will be Doppler shifted, we use the mapping from \cite{Brogi2019} as described by,
\begin{equation} \label{eq: logL}
\log(L) = -\frac{N}{2}\log[s_{f}^{2} - 2R(l) + s_{g}^{2}], 
\end{equation}
where the cross-covariance, $R(l)$, is related to the correlation coefficient by,
\begin{equation}
    C(l) = \frac{R(l)}{\sqrt{s_{f}^{2} s_{g}^{2}}}.
    \label{eq: correlation coefficient}
\end{equation}
In equation~\ref{eq: logL}, $N$ is the number of wavelength channels used in the cross-correlation and $s_{f}^{2}$ and $s_{g}^{2}$ refer to the variance of the data and the model, respectively. 
In equations~\ref{eq: logL} and \ref{eq: correlation coefficient}, $l$ represents the cross-correlation lag, that is equal to the planet's Doppler shift at the time the $\log(L)$ is evaluated. Using this form of the $\log(L)$, we are assuming the case where an additional atmospheric scaling factor $a$ is equal to unity. This is necessary because we treat the scaling factor as a model parameter, and therefore we apply it to the model spectrum {\sl prior} to the likelihood computation (Equation~\ref{equation: scaled flux}), which is necessary to account for the effects of the analysis on the model. Using this method, we are able to directly convert the correlation values into a log likelihood velocity map.

The log-likelihood values from Equation~\ref{eq: logL} are calculated for each night, each order and each spectrum and subsequently summed to obtain a single log-likelihood value for each model and each set of parameters,
\begin{equation} \label{eq: total_logL}
\log(L)_\mathrm{tot} = -\sum_{k=1}^4 \sum_{i=1}^{N_\mathrm{o}} \sum_{j=1}^{N_{\text{s}}} \frac{N_{kij}}{2} \log[s_{f,kij}^{2} - 2R_{kij}(l) + s_{g,kij}^{2}],
\end{equation}
where the index $k$ denotes the observing night, $i$ the spectral order, and $j$ the frame number. We further note that the number of orders $N_{\mathrm{o}}$ and the number of spectra $N_{\text{s}}$ vary from night to night.

\subsection{Atmospheric models} \label{subsec: models}

\begin{table*}
    \centering
    \caption{The grid of models that were used in the analysis. The range of abundances tested for each species are shown, these varied in steps of 1 dex. For each species, the abundance was also allowed to drop to zero in each opacity grid to model each species individually. The exceptions to this are the water abundances which were fixed to $10^{-3.3}$ and $10^{-3}$ for the water only models (top row) and the H$_{2}$O, HCN and C$_{2}$H$_{2}$ grid (bottom row), respectively.}
    \begin{tabular}{lcccc}
    \hline
    \textbf{Opacity source(s)} & $\log_{10}(\mathrm{VMR})$ & $T_{1}$ (K) & $T_{2}$ (K) & \textbf{Number of models}\\
    \hline
    $\mathrm{H_{2}O}$ & $\text{-}3.3$ & [1400,1600,1800,2000] & [800,1200,1600,2000] & 12 \\
    $\mathrm{H_{2}O}$, $\mathrm{CH_{4}}$, HCN & $\text{-}3.0\,\mathrm{to}\,\text{-}5.0$, $\text{-}4.0\,\mathrm{to}\,\text{-}6.0$, $\text{-}5.0\,\mathrm{to}\,\text{-}7.0$ & 1800 & 1200 & 64 \\
    $\mathrm{H_{2}O}$, $\mathrm{NH_{3}}$, $\mathrm{C_{2}H_{2}}$ & $\text{-}3.0\,\mathrm{to}\,\text{-}5.0$, $\text{-}4.0\,\mathrm{to}\,\text{-}6.0$, $\text{-}5.0\,\mathrm{to}\,\text{-}7.0$ & 1800 & 1200 & 64 \\
    $\mathrm{H_{2}O}$, HCN, $\mathrm{C_{2}H_{2}}$ & $\text{-}3.0, \text{-}2.0\,\mathrm{to}\,\text{-}7.0, \text{-}2.0\,\mathrm{to}\,\text{-}7.0$ & 1800 & 1200 & 49 \\
    \hline
    \end{tabular}
    \label{tab: model grids}
\end{table*}

To model the thermal emission from \tauboo b, we use the line-by-line radiative transfer code \textsc{genesis} \citep{Gandhi2017}. These model spectra were produced using the same methods as described in \cite{Hawker2018}, \cite{Cabot2019}, \cite{Webb2020} and \cite{Gandhi2020}. The spectra were generated at a wavelength range of $\lambda = $ 0.96\,-\,1.8\mum, with a wavenumber spacing of 0.01\,cm$^{-1}$ which corresponds to a resolution of $R(\lambda$) = $\frac{10^{6}}{\lambda}$, where $\lambda$ is in \mum. Before cross-correlation, these models were re-grid to a constant resolution (i.e.  $\lambda/\Delta\lambda$) and then convolved with a Gaussian kernel with a FWHM equivalent to the instrumental resolution of CARMENES in the NIR ($R=80,400$). We also assume each model spectrum has a 1D $T-p$ profile that has been parameterised by upper ($T_{2}, p_{2}$) and lower ($T_{1}, p_{1}$) points in the atmosphere. Above ($p < p_{2}$) and below ($p > p_{1}$) these points the atmosphere is assumed to be isothermal.

Opacity sources were included from the following line list databases; HITEMP, for H$_{2}$O \citep{Rothman2010} and CH$_{4}$ \citep{Hargreaves2020}, ExoMol for HCN \citep{Harris2006,Barber2014}, NH$_{3}$ \citep{Coles2019} and C$_{2}$H$_{2}$ \citep{Chubb2020}. We note that CARMENES has no sensitivity to CO in the \textit{Y}, \textit{J} and \textit{H}-bands and therefore we do not include this molecule in the modelling. Even though there are CO lines at $\sim1.6$\mum, these are over two orders of magnitude weaker than in the \textit{K}-band \citep[e.g.][]{Gandhi2020}. In addition, at $1.6$\mum the water opacity is stronger than for CO, thus the weaker CO lines will be significantly shielded by water lines. These models also include collisionally induced absorption from H$_{2}$-H$_{2}$ and H$_{2}$-He \citep{Rothman2012} and broadening from each opacity source \citep{Gandhi2020a}. In table~\ref{tab: model grids} we show the grids of models that were used in our analysis tested against the observed spectra. We fix the pressure points to be $p_{1} = 1$ and $p_{2} = 10^{-3}$ bars for all the models in the grids. For pressures lower that $10^{-3}$\,bars, the $T-p$ profile is predicted to be isothermal from GCM and 1D modelling \citep[e.g.][]{Beltz2021}, therefore, most of the core of the lines will be formed at pressures of $>10^{-3}$\, bars.
We generated a $T-p$ profile grid to explore various temperature gradients for the atmosphere with a single opacity source of water with a fixed chemical equilibrium and solar composition abundance of VMR$=10^{-3.3}$. The choice of temperatures was guided by the range of equilibrium temperatures of the day-side of the planet, $T_{\mathrm{eq}} = 1600-2000$\,K, which depends on the efficiency of day-night heat redistribution. We also generated a large opacity grid with water combined with further minor species at varying abundances, including at zero abundance. We assume that the abundance for each opacity source in all of the models have a constant VMR with pressure. For these opacity grids, we fixed the temperature gradient to $T_{1} = 1800$ and $T_{2} = 1200$\,K. We note here that with a fixed grid of models, we are limited in our ability to constrain the atmospheric lapse rate and the molecular abundances individually due to the partial correlation between these parameters. A full atmospheric Bayesian retrieval with free parameters for the T-p profile and abundances will constrain these parameters individually for which we defer to a future study.


As with the data, we also pass these models through the telluric removal pipeline as described in section~\ref{sec: tell removal} prior to cross-correlation. This is to replicate the unavoidable scaling effects that occur to the atmospheric signal in the telluric removal sequence, thus avoiding potential biases in the cross-correlation analysis (see final panel in Fig. 2 of \cite{Brogi2019}.

\section{Results} \label{sec: results}

\subsection{Single species analysis} \label{subsec: single species analysis}

\begin{figure}
    \centering
    \includegraphics[width=\columnwidth]{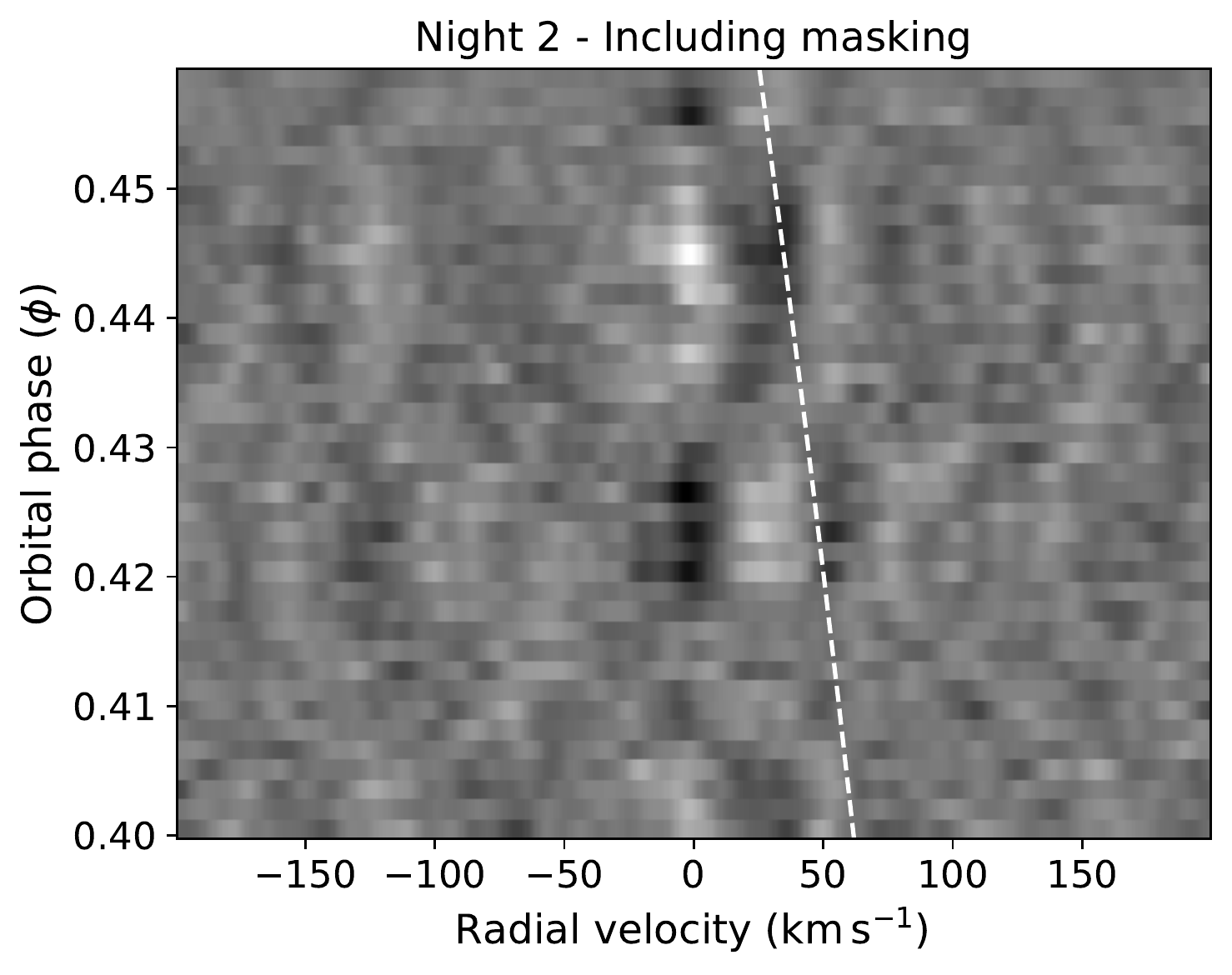}
    \caption{The relative CC-$\log(L)$ values binned in phase with a pure water model with a $\log(\mathrm{VMR})=-3.0$ with the spectra from night 2. The spectra have been co-added in $\phi=0.0015$ wide bins. Darker shades indicate anti-correlation whereas lighter shades indicate correlation with the water model. The white dashed line shows the expected radial velocity of \tauboo b. The strong correlation and anti-correlation stripe in the telluric rest-frame (i.e. $\sim0\,$km\,s$^{-1}$) suggests strong telluric absorption in these spectra.}
    \label{fig: n5 logl ph bin}
\end{figure}

\begin{figure}
    \centering
    \includegraphics[width=\columnwidth]{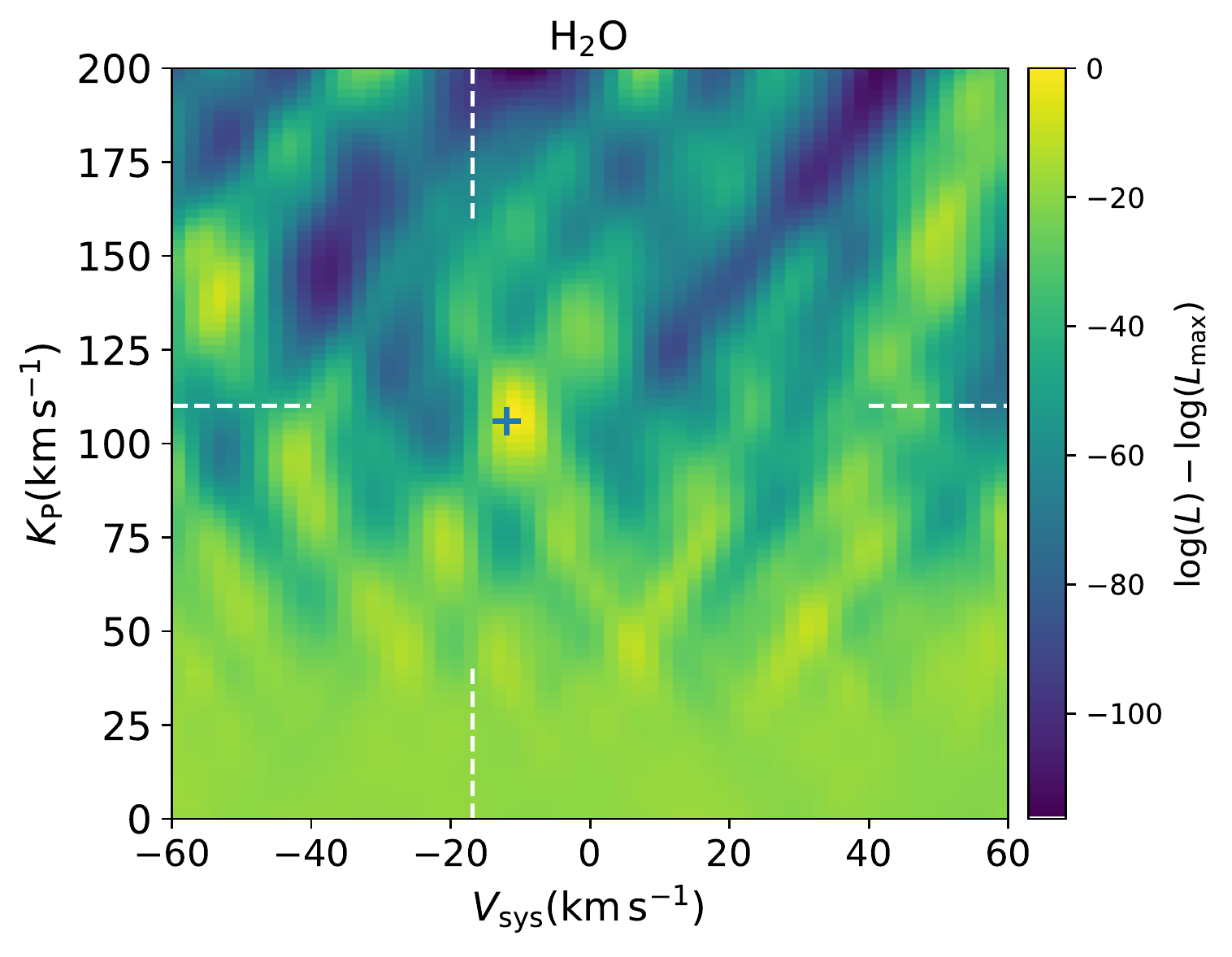}
    \caption{$K_{\mathrm{P}}$-$V_{\mathrm{sys}}$ velocity map of the CC-$\log(L)$ mapping of the observed spectra with the best-fitting water model with a VMR$=10^{-3}$ in $\Delta\log(L)=\log(L)-\log(L)_{\mathrm{max}}$. The white dashed lines indicate the expected position of the signal from \tauboo b from the literature. The blue cross shows the location of the $\log(L_{\mathrm{max}})$.}
    \label{fig: best fit signal from 4 nights}
\end{figure}

\begin{figure*}
    \centering
    \includegraphics[width=\textwidth]{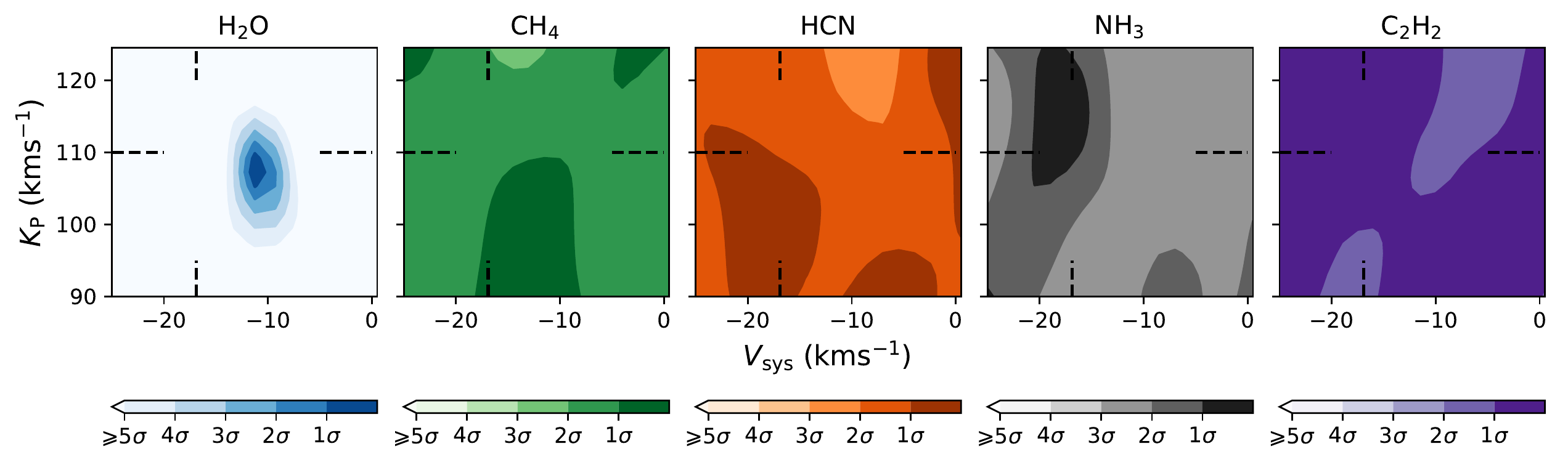}
    \caption{CC-$\log(L)$ significance $K_{\mathrm{P}}$ - $V_{\mathrm{sys}}$ maps of the five species at solar abundance co-added for the four nights of spectra. The filled contours indicate the areas of significance away from the peak in the $\log(L)$. From left to right, we show the maps of; H$_{2}$O, CH$_{4}$, HCN, NH$_{3}$ and C$_{2}$H$_{2}$. The H$_{2}$O map shows a zoomed in version of the signal seen in  Fig.~\ref{fig: best fit signal from 4 nights}. The black dashed lines indicate the location of the orbital solution given from the literature. There is a clear signal from water close to the velocity of the planet and no evidence for any other minor species.}
    \label{fig: single species}
\end{figure*}

We first analysed the data against five single species individually at abundances of $\log_{10}$VMR; H$_{2}$O = $-3.0$ to $-5.0$, CH$_{4}$ = $-4.0$ to $-6.0$, HCN = $-5.0$ to $-7.0$, NH$_{3}$ = $-4.0$ to $-6.0$ and C$_{2}$H$_{2}$ = $-5.0$ to $-7.0$. These tests were done on the combined species models on rows 2 and 3 in table~\ref{tab: model grids} with the additional species effectively removed by setting their abundances to zero. For H$_{2}$O, the range of abundances used corresponds to the expected solar abundance assuming a solar C/O ratio and metallicity at thermochemical equilibrium down to sub-solar values approaching the upper-limit determined in \cite{Pelletier2021} ($\log_{10}(\mathrm{VMR})<-5.66$). For the remaining species, the expected solar abundances are expected to be too low to be observable for a planet with a temperature of $\sim1800\,$K at a solar C/O ratio and metallicity \citep[e.g.][]{Madhusudhan2012, Moses2013}. However, we tested a range of enhanced abundances for CH$_{4}$, HCN, NH$_{3}$ and C$_{2}$H$_{2}$ to include potential scenarios where the atmospheric C/O ratio and metallicity are super-solar which has been observed in recent high resolution atmospheric studies \citep{Giacobbe2021}. For these modelling tests, we do not optimise the atmospheric scaling parameter, i.e. $a=1$.      

In Fig.~\ref{fig: n5 logl ph bin}, we show the result of correlating the reduced spectra from night 2 with a pure water model for \tauboo b, shifted in radial velocity. The prominent alternating pattern of correlation (lighter shades) and anti-correlation features (darker shades) at zero lag radial velocity shows that the tellurics are strongly correlating with the atmospheric water models. Even though the radial velocity trail of \tauboo b is significantly shifted from the telluric rest frame, the correlated telluric noise is overwhelming any potential signal from the planet. Therefore, the inclusion of these spectra will cause spurious telluric noise in the CC-$\log(L)$ analysis and have subsequently been removed from the rest of the analysis (see appendix~\ref{app: CC to logl all nights} and Fig.~\ref{fig: h2o logl all nights }). We show the same phase resolved correlation with a water model with all nights combined in Fig.~\ref{fig: all nights logl ph bin} in appendix~\ref{app: CC to logl all nights}. 

For the four nights of spectra, co-added in time and wavelength, we observe a signal for the presence of water vapour in the thermal emission of this atmosphere which is shown in Fig.~\ref{fig: best fit signal from 4 nights}. We find that the water signal peaks at a planet velocity semi-amplitude of $K_{\mathrm{P}} = 106.0^{+2.8}_{-2.2}$\kms which is consistent within $1\,\sigma$ from the literature values \citep{Brogi2012,Lockwood2014,Pelletier2021}. However, we find a shift in the systemic velocity of \tauboo with a $V_{\mathrm{sys}} = -12.0^{+1.0}_{-0.5}$\kms which is a $\sim$+5.4\kms shift from the expected $V_{\mathrm{sys}} = -16.9$\kms \citep{GaiaDR2,Justesen2019}.

To test for the preference in the water abundance in the modelling, we set the abundances for the combined species in the second row of table~\ref{tab: model grids} to zero, i.e. CH$_{4}$ and HCN are set to zero but H$_{2}$O is allowed to vary. We find that an abundance of VMR$_{\mathrm{H_{2}O}} = 10^{-3}$ is strongly preferred over the lower abundances of $10^{-4}$ and $10^{-5}$ by $3.8\,\sigma$ and $5.0\,\sigma$, respectively. In Fig.~\ref{fig: single species}, we show the confidence intervals in the $K_{\mathrm{P}}-V_{\mathrm{sys}}$ correlation maps for each individual species. This shows the tight confidence intervals from the water signal that we observe in the continuous CC-$\log(L)$ map in Fig.~\ref{fig: best fit signal from 4 nights} and no evidence for the presence of other minor species in this analysis. To calculate confidence intervals, we converted the CC-$\log(L)$ mapping to a chi-square distribution (with two degrees of freedom, $K_{\mathrm{P}}$ and $V_{\mathrm{sys}}$) from the peak in the log-likelihood, $\log(L_{\mathrm{max}})$, using Wilks' theorem \citep{wilks1938}, $\chi^{2} = -2\Delta\log(L) = -2\log(L/L_{\mathrm{max}})$. Hence, we can determine the p-values from $\log(L_{\mathrm{max}})$ by halving the two-tail survival function of the $\chi^{2}$ distribution. Finally, we are able to convert these p-values into $\sigma$ levels by calculating the normal distribution inverse survival function.

We also analysed the water signal against varying $T-p$ profiles (see section~\ref{subsec: models}) with a fixed  water abundance of $\mathrm{VMR}=10^{-3.3}$. We find that a steeper temperature gradient is preferred for this atmosphere with $T_{1} = 1800$\,K preferred over lower temperatures by 3.2\,$\sigma$. With $T_{1}$ fixed to 1800\,K, the upper temperature $T_{2} = $ 800\,K is marginally preferred over higher temperatures by $1.6\,\sigma$ (1200\,K) and $2.9\,\sigma$ (1600\,K). We do note however that these are only qualitative constraints on these parameters as the atmospheric lapse rate and chemical abundances partially correlate. Recent studies with high resolution spectroscopy observations \citep{Pelletier2021, Line2021} have shown that by using a full Bayesian atmospheric retrieval, the lapse rate and the absolute abundances can be retrieved will little correlation between these parameters. Thus, constraining both parameters in these spectra will be possible with a full atmospheric retrieval analysis for which we defer to a future study.

\subsection{Combined species analysis} \label{subsec: combined species}

\begin{figure}
    \centering
    \includegraphics[width=\columnwidth]{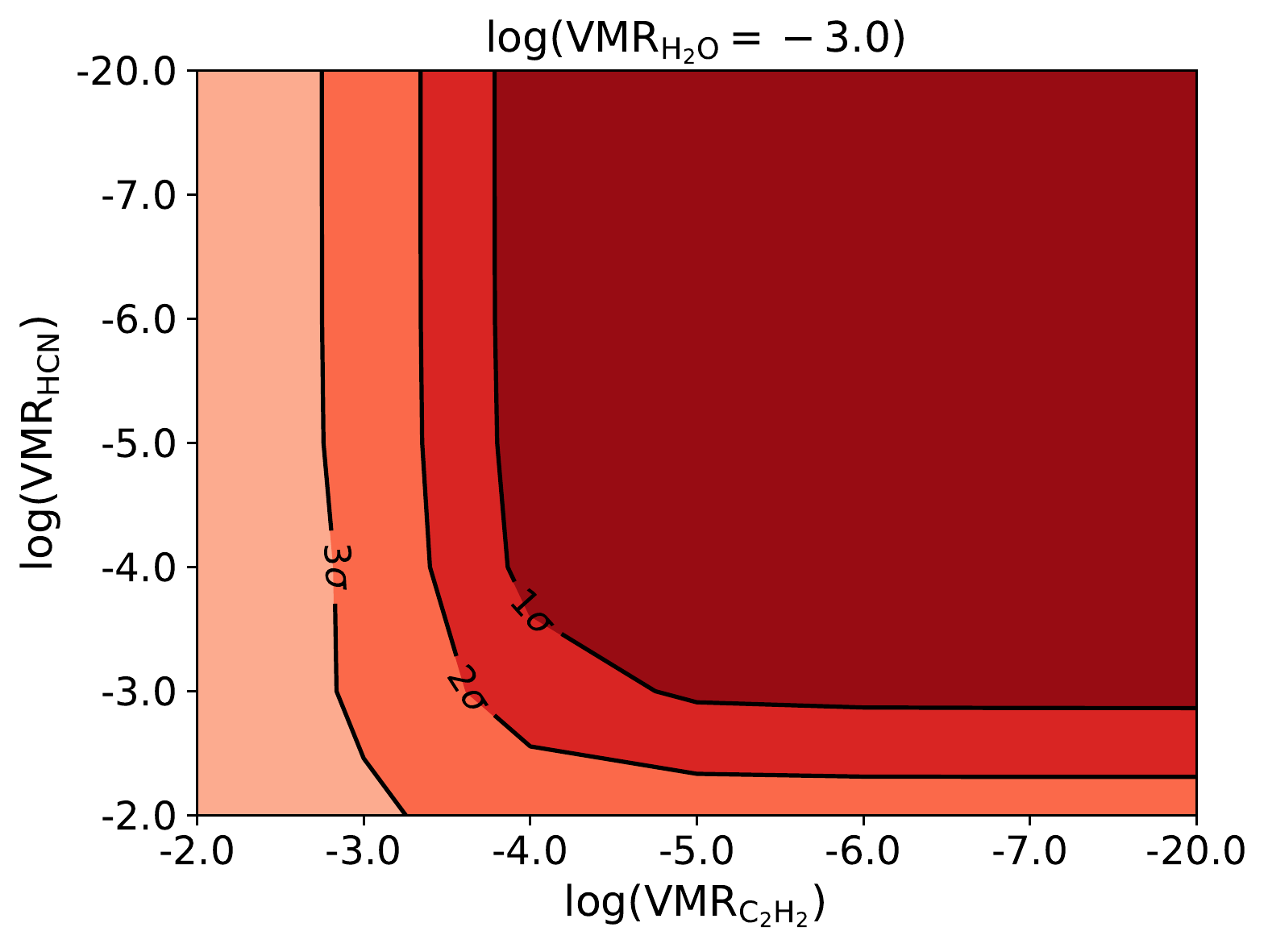}
    \caption{Abundance constraints on combining the best-fitting water abundance with additional HCN and C$_{2}$H$_{2}$ or without (abundance of -20.0). There is no preference for the addition of these additional species and we can only place upper limits on the abundances based on this analysis.}
    \label{fig: abundance constraints 4 nights}
\end{figure}

In addition to the individual species, we also analysed a grid of models that combined the additional minor species to the best-fitting water model ($\log(\mathrm{VMR})=-3.0$). Initial tests with models that have solar abundances in chemical equilibrium, we saw a marginal increase ($<1\,\sigma$) in the $\log(L_{\mathrm{max}})$ with the addition of HCN and C$_{2}$H$_{2}$ in the modelling. Therefore, we expanded the abundance range of these species with water fixed at VMR$=10^{-3}$ to super-solar values (see the bottom row in table~\ref{tab: model grids}) to explore whether these species significantly increase the detection significance from the pure water models. In Fig.~\ref{fig: abundance constraints 4 nights}, we show the abundance constraints on the grid containing HCN and C$_{2}$H$_{2}$. We find no evidence $>1\,\sigma$ that these additional  species in the modelling improve the detection significantly from the pure water models in this analysis. However, we can place a $3\,\sigma$ and $2\,\sigma$ upper limits on the abundance of C$_{2}$H$_{2}$ and HCN at a VMR$=10^{-3}$ and VMR$=10^{-2.5}$, respectively. Again, we note that these limits in abundance are dependent on the choices for the change of temperature with pressure, i.e. the atmospheric lapse rate. 



\subsection{MCMC analysis} \label{subsec: mcmc}

\begin{figure}
    \centering
    \includegraphics[width=\columnwidth]{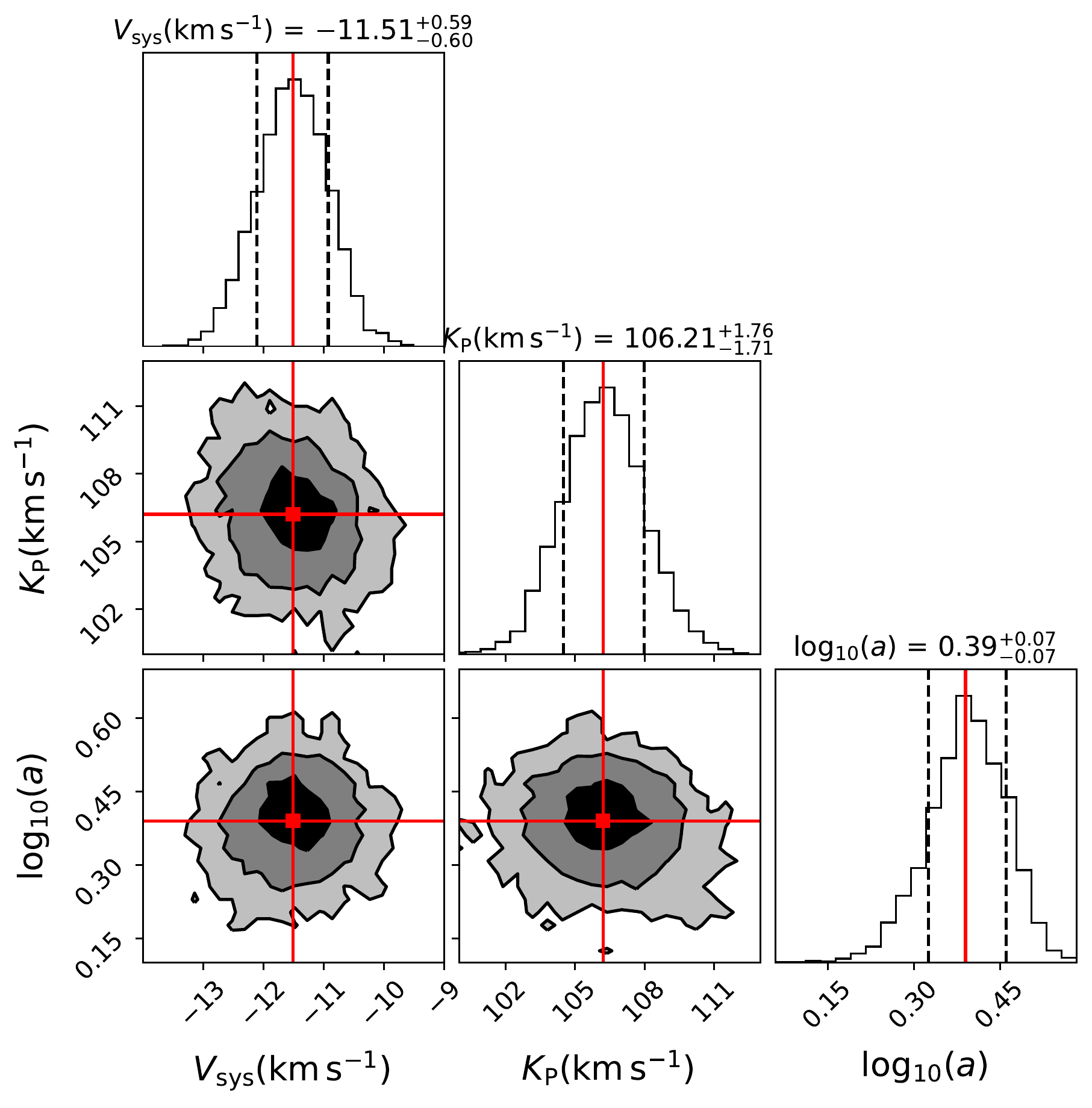}
    \caption{Posterior distributions of the orbital semi-amplitude ($K_{\mathrm{P}}$), the systemic velocity ($V_{\mathrm{sys}}$) and the logarithm of the atmospheric scaling parameter ($\log_{10}(a)$) retrieved from the best-fitting model. The median values for each parameter are given by the solid red lines on the corner plots and histograms. The black dashed lines on the histograms show the 0.16 and 0.84 quantiles. The filled in contours show the 1, 2 and 3 $\sigma$ regions (darkest to lightest shades, respectively). The retrieved posteriors shows a constructive, co-added signal is retrieved from the best-fitting atmospheric model}
    \label{fig: 4 nights corner}
\end{figure}

\begin{figure}
    \centering
    \includegraphics[width=\columnwidth]{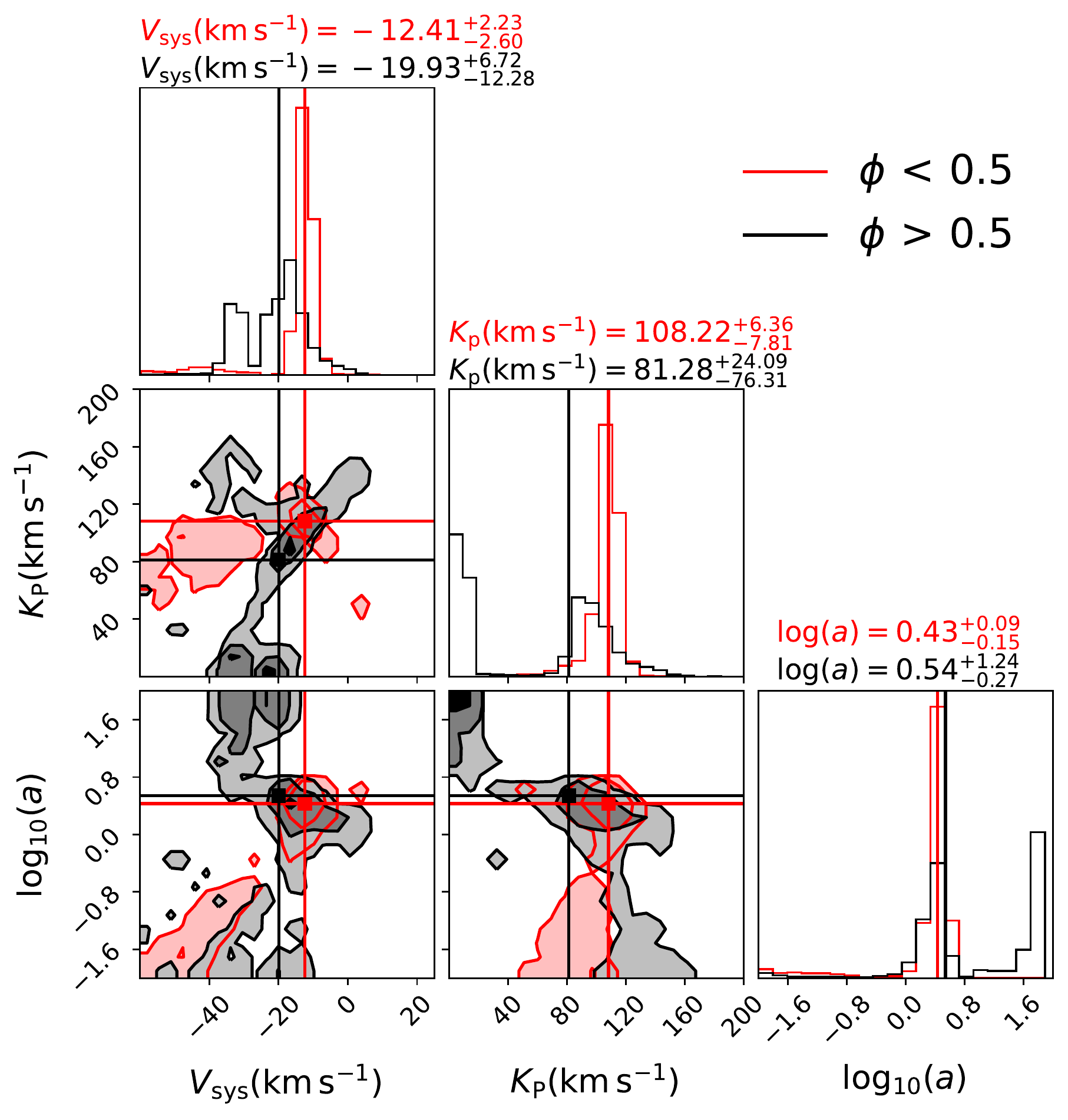}
    \caption{Same as Fig.~\ref{fig: 4 nights corner} but the data-set has been split into pre- (red) and post-superior (black) conjunction. The red and black solid lines show the median values of the pre- and post-superior conjunction data-set, respectively.}
    \label{fig: pre&post corner}
\end{figure}

We performed an MCMC analysis on this data-set with the best-fitting model from the opacity model grid, i.e. a pure water model with a VMR=$10^{-3}$ and a T-p profile of ($T_{1}, p_{1}$) = (1800\,K, 1\,bar) and ($T_{2}, p_{2}$) = (1200\,K, $10^{-3}$\,bars). For the combined four nights data-set of 560 spectra, it was too computationally expensive ($\sim82$\,s per step pooled over 40 processes) to viably explore all of the model grid and thus only the best-fitting model was used here. For this analysis, we used the \texttt{emcee} \citep{emcee} package with 12 walkers and a chain-length of 1000, resulting in a total chain length of 12,000 points. We fit for a two parameter circular orbital solution ($K_{\mathrm{P}}$ and $V_{\mathrm{sys}}$) and an atmospheric scaling parameter ($\log_{10}(a)$) which allows the line strengths to vary. We allowed the MCMC to sample from uniform distributions of the prior parameters with ranges of, $-60<V_{\mathrm{sys}}\,(\mathrm{km\,s^{-1}})<60$, $0<K_{\mathrm{P}}\,(\mathrm{km\,s^{-1}})<200$ and $-2.0<\log_{10}(a)<2.0$. We initialise each Markov chain with the parameters set to the expected literature values, i.e. $K_{\mathrm{P}} = 110$\kms, $V_{\mathrm{sys}} = -16$\kms and $\log_{10}(a) = 0$. The evolution of each chain is also driven by the CC-$\log(L)$ mapping as described in section~\ref{subsec: CC and logL mapping}.

\begin{table}
\centering
\caption{Retrieved posterior values of \tauboo b from a MCMC analysis with the best fitting atmospheric model.}
\begin{tabular}{lccc}
\hline
\textbf{Data-set} & \textbf{$V_{\mathrm{sys}}$(\kms)} & \textbf{$K_{\mathrm{P}}$(\kms)} & $\log_{10}(a)$\\
\hline
$\phi\,<\,0.5$ & $-12.41^{+2.23}_{-2.60}$ & $108.22^{+6.36}_{-7.81}$ & $0.43^{+0.09}_{-0.15}$ \\
\rule{0pt}{3ex}$\phi\,>\,0.5$ & $-19.93^{+6.72}_{-12.28}$ & $81.28^{+24.09}_{-76.31}$ & $0.54^{+1.24}_{-0.27}$ \\
\rule{0pt}{3ex}Combined & $-11.51^{+0.59}_{-0.60}$ & $106.21^{+1.76}_{-1.71}$ & $0.39\pm0.07$ \\
\hline
\end{tabular}
\label{table: posteriors}
\end{table}

In Fig.~\ref{fig: 4 nights corner}, we show the retrieved posterior distributions from the combined set of spectra. It is clear that all of the spectra do co-add constructively to converge onto a single orbital solution from the water detection with $K_{\mathrm{P}} = 106.21^{+1.76}_{-1.71}$ and $V_{\mathrm{sys}} = -11.51^{+0.59}_{-0.60}$. The retrieved scaling factor of $\log_{10}(a) = 0.39$ suggests that the observed water lines are $\sim2.5\times$ deeper compared to the best-fitting model. For these retrieved parameters, we determine a detection significance of $5.5\,\sigma$ when comparing the $\log(L)$ values of the best-fitting water model and a featureless spectrum (i.e. a blackbody spectrum). 


To test whether there is any phase dependence on the retrieved atmospheric parameters, we have split the data into pre- (i.e. all spectra $\phi<0.5$, i.e. night 4 only) and post the superior conjunction (all spectra $\phi\,>0.5$). For our data-set with only the four nights used, the pre- and post-superior conjunction spectra includes a phase coverage of $\phi = 0.420-0.481$ and 0.511-0.599, respectively. In Fig.~\ref{fig: pre&post corner}, we show a corner plot of the posteriors from the MCMC using only the pre-superior conjunction data in red, and the post-superior conjunction data in black. In table~\ref{table: posteriors}, we show the retrieved parameters from the two data-sets. It can be seen that the spectra at pre-superior conjunction provides a much more convincing detection of water with tighter constraints retrieved from the posterior distributions which gives a detection significance of water at $4.6\,\sigma$ for night 4 alone. The retrieved posteriors for the two data-sets do, however, show some overlap at the $2\,\sigma$ level at the expected radial velocity of \tauboo b suggesting a weak agreement between the two data-sets. However, neither data-set converges to a single solution with the post-superior conjunction posteriors in particular showing a double peak in the distribution. This indicates that the detection of water is weak in the individual nights and that we need to co-add the full data-set of 560 spectra in order to converge onto a single solution from the water detection. The failure of convergence for both these data-sets means that we are unable to constrain the individual atmospheric scaling factors, therefore, we find no evidence for a phase dependence on the observed water signature from \tauboo b.


\section{Discussion} \label{sec: discussion}

Using the CARMENES high resolution instrument, we detect a signature of water in absorption in the day-side spectrum from the non-transiting planet \tauboo b. Co-adding all five nights of spectra, we determine an orbital solution with a $K_{\mathrm{P}} = 106.21^{+1.76}_{-1.71}$\kms and a $V_{\mathrm{sys}} = -11.51^{+0.59}_{-0.60}$\kms. Using a stellar mass of $1.35\pm0.03\,\mathrm{M_{\odot}}$ \citep{Takeda2007} and a radial velocity of $468.42\pm2.09\,\mathrm{m\,s^{-1}}$ \citep{Justesen2019}, we derive a planetary mass of $M_{\mathrm{P}} = 6.24^{+0.17}_{-0.18}$\,$\mathrm{M_{J}}$. Furthermore, we are able to derive an inclination of $i = 41.6^{+1.0}_{-0.9}$\,degrees determined from a planet separation of $0.04869^{+0.00039}_{-0.00040}$\,AU \citep{Rosenthal2021}, which is consistent with those determined in \cite{Lockwood2014} and \cite{Pelletier2021}. Despite having $\sim3\times$ weaker constraints on $K_{\mathrm{P}}$, we are able to match the constraints on the inclination and the mass of the planet due to the recent improvement on the constraints on the measured semimajor axis \citep{Rosenthal2021} than what was used in \cite{Pelletier2021}. This highlights the importance of regular improvements to the system parameters on planets with large monitoring radial velocity surveys such as in \cite{Rosenthal2021} which are essential for retrieving precise orbital parameters in high resolution atmospheric studies.

We find that the best fitting model requires a water abundance of VMR = $10^{-3}$, which is consistent with solar, and a non-inverted $T-p$ profile of 1800-1200\,K over a pressure range of 1-$10^{-3}$\,bars. We find no evidence for the presence of any further minor species from this analysis. We also split the spectra into a pre- and post-superior conjunction and we find that there is only weak evidence for a detection of water in the post-dayside spectra, this is likely due to the poorer observing conditions for those nights (see section~\ref{sec: Observations}) and perhaps an overall weak signal from the atmosphere. This may also hint at a phase dependence on the water signal from \tauboo b, however, we find no evidence for this in these spectra. For this model, we also retrieve a scaling factor of $\log_{10}(a) = 0.39$ indicating that the water depths are $\sim2.5\times$ deeper that the modelled spectral lines. As this planet is at an $\sim41.6^{\circ}$ inclination, we are viewing a mixture of the hot day-side and the cooler night-side of the atmosphere. Due to the absence of external irradiation from the host star on the night-side, it is expected that the lapse rate is steeper compared the day-side resulting in steeper spectral lines \citep[e.g.][]{deKok2014}. As water is expected to be formed on both sides \citep{Madhusudhan2012}, it is likely that our scaling factor is trying to compensate for the differences in line depths over the two temperature regions.

\subsection{The orbital parameters of non-transiting planets}

\begin{figure}
    \centering
    \includegraphics[width=\columnwidth]{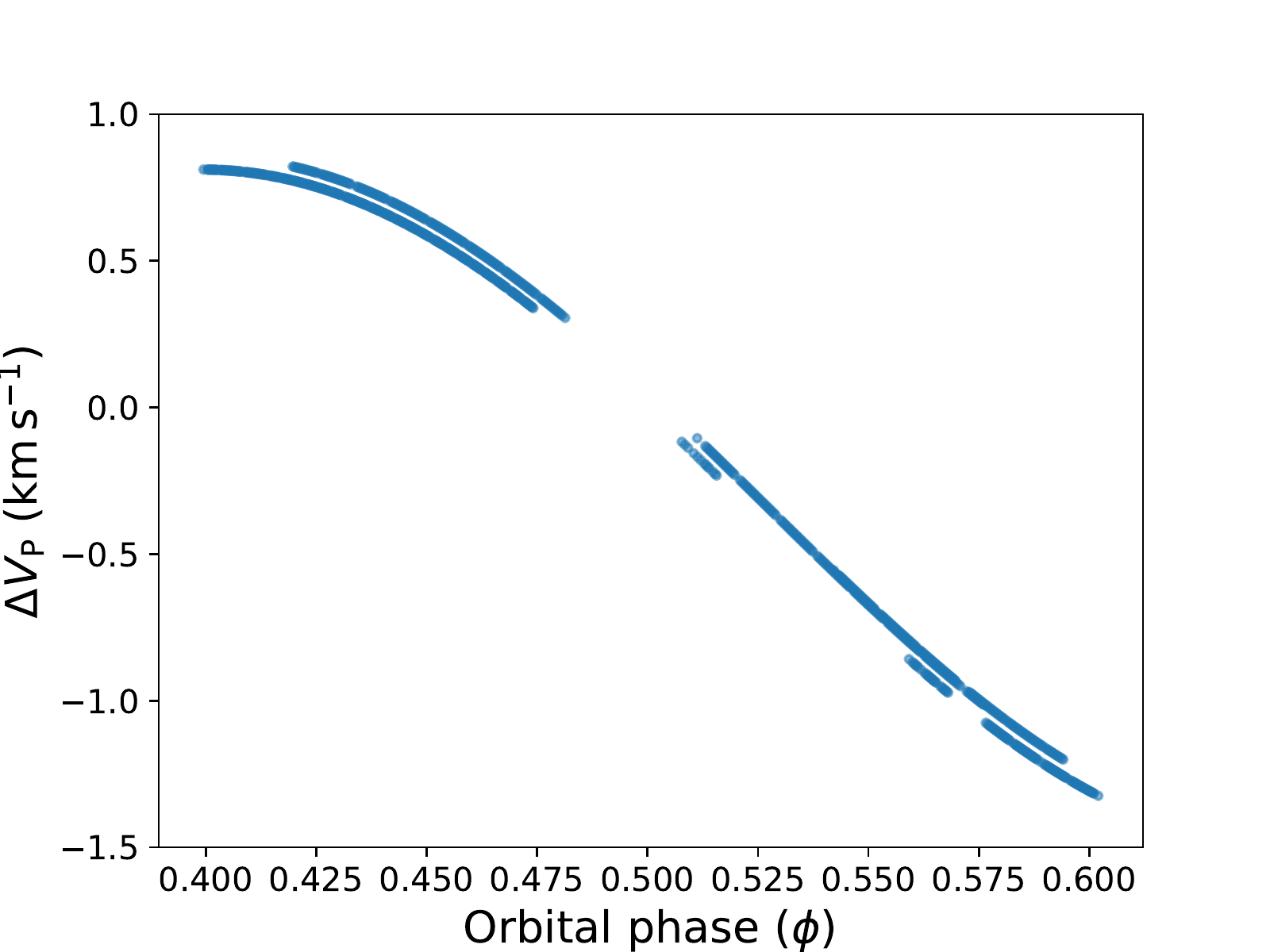}
    \caption{Difference in the radial velocity of \tauboo b in \kms between the eccentric and circular orbital solutions. The magnitude of these velocity shifts do not explain the $\sim7$\kms shift we retrieve from the systemic velocity.}
    \label{fig: deltaRV from ecc}
\end{figure}

Assuming a circular orbit for \tauboo b, we retrieve a systemic velocity of $-11.51^{+0.59}_{-0.60}$\kms which is significantly shifted from the literature value of -16.9$\pm$0.3\kms \citep{GaiaDR2}. However, these observations were taken a few years apart from those from the Gaia data release 2 survey, therefore, according to the radial velocity solution of \tauboo A in \cite{Justesen2019}, it is likely that the systemic velocity has shifted further to over $-17$\kms. This is due to the systems M-dwarf companion, \tauboo B, approaching periastron which will impart an additional radial velocity shift of \tauboo A from the planetary signal (see Fig.~3 in \cite{Justesen2019}). The analysis by \cite{Pelletier2021} also recovers a significantly shifted velocity of $V_{\mathrm{sys}} = -15.4\pm0.2$\kms. This shift in our $V_{\mathrm{sys}}$ can be partially explained by assuming an eccentric solution from \cite{Borsa2015} ($e = 0.011\pm0.006$ and $\omega = 113.4^{\circ}\pm32.2^{\circ}$), as used in \cite{Pelletier2021}. If we used this eccentric solution on the phases calculated using the updated $T_{0}$ and the orbital period from \cite{Justesen2019}, this would result in a shift in the velocity by $\sim-3$\kms. If we do adopt the full orbital solution from \cite{Borsa2015} (including their retrieved $T_{0}$ and orbital period), then we get at most a shift in the planetary velocity by $\pm1.5$\kms as show in Fig.~\ref{fig: deltaRV from ecc} also does not resolve the discrepancy in the retrieved systemic velocity. However, the time of periastron ($T_{0}$) for the eccentric solution obtained in \cite{Borsa2015} is highly uncertain ($\pm0.3$ in BJD) and could therefore result in an even greater shift of several \kms for the systemic velocity. Due to this uncertainty in time of periastron from the eccentric orbital solution, we do not implement this solution into our analysis and instead adopted the circular orbital solution from \cite{Justesen2019}. However, this does show that even a relatively small eccentricity can lead to significantly shifted planetary velocities of up to several \kms if the eccentric solution is highly uncertain. This was also true in the high resolution characterisation of the non-transiting planet 51 Pegasi b \citep{Birkby2017} which needed to invoke a large shift in the time of periastron by $\Delta T_{0}=0.07$ in days for the circular orbital solution to match the observed water signal to that of the observed systemic velocity. This highlights the importance of retrieving and regularly updating precise orbital solutions with long period radial velocity surveys in order to accurately constrain the planetary velocities with follow-up high resolution atmospheric characterisation, particularly for non-transiting systems. Further observations of \tauboo b are therefore necessary to try explain these apparent discrepancies in the characterisation of this system.

\subsection{Comparison with previous analyses of the atmosphere of \tauboo b} \label{subsec: disc on previous analyses}

Our analysis confirms the detection of water in the \textit{L}-band from \cite{Lockwood2014} who used the NIRSPEC instrument at the Keck Observatory. Curiously, our water detection is in stark contrast to the results in \cite{Pelletier2021} who find only a $3\,\sigma$ upper limit on the water abundance at a VMR$ = 10^{-5.66}$ with a full atmospheric retrieval. This analysis detects the presence of water at a VMR$= 10^{-3}$ which is preferred over a VMR of $10^{-4}$ and $10^{-5}$ by $3.8\,\sigma$ and $5.0\,\sigma$, respectively, with a non-inverted $T-p$ profile of 1800-1200\,K over $1-10^{-3}$\,bars. However, we do emphasise that the molecular abundance and $T-p$ profile in our analysis will be partially correlated and therefore a full atmospheric retrieval is needed on these spectra to give an accurate comparison of the water abundance with \cite{Pelletier2021}. 

The analysis by \cite{Pelletier2021} observed  the day-side thermal emission of \tauboo with the SPIRou ($R=70,000$) instrument over five nights of data spanning a similar phase coverage to this analysis. Due to the wider wavelength coverage of SPIRou ($\lambda=0.95-2.50$\,\mum) compared with CARMENES, it was expected that SPIRou should have observed the day-side emission of CO and H$_{2}$O simultaneously, however, only CO at was detected in these spectra. As CARMENES does not cover the strong 2-0 R-branch absorption feature at $\sim$2.3\mum, we cannot observe these two molecules simultaneously. \cite{Brogi2012} also only detected the absorption features from CO from \tauboo b, however, their observations were taken with a narrow wavelength range to cover the 2.3\mum feature, therefore, it is likely that the strong CO lines obscured the H$_{2}$O lines to be observable. It could also be the case that \cite{Pelletier2021} suffer from the same sort of behaviour as the strong CO lines mask the weaker water absorption features in the atmosphere of \tauboo b, although this explanation is perhaps unsatisfactory as they deduce that the spectra are sensitive to the presence of water from \tauboo b with their injection and retrieval tests. We are only able to detect a convincing signal from water our fourth night which has a phase coverage of $\sim$0.42-0.48. The corresponding night in \cite{Pelletier2021} that covered the same phase range suffered from poorer seeing and a slightly lower SNR than the rest of their nights. If it is the case that there is some phase dependence and the signal is far stronger within this phase range, then this could explain the absence of water in their analysis. However, if there is a constant abundance of water across the orbit then it should be observable in the spectra from \cite{Pelletier2021}. The final explanation for the discrepancies in the water detection could be due to differing line lists used in each analysis. In this analysis we use the HITEMP water opacities calculated using the BT2 line list \citep{Barber2006} as also used in the analysis \cite{Pelletier2021}, therefore, we rule out the possibility that line lists are the cause of the discrepancies between our two analyses. 

\section{Conclusions} \label{sec: conclusions}

With $\sim21$\,h of observations over five nights at high resolution with the CARMENES spectrograph, we unambiguously detect the presence of absorption features from water vapour through thermal emission from the atmosphere of \tauboo b. We searched for, HCN, CH$_{4}$, NH$_{3}$ and C$_{2}$H$_{2}$ but found no evidence for these minor species. Using a grid of models, we found that these spectra prefer a high abundance of water (VMR = $10^{-3}$) which is significantly preferred over models with lower abundances by $\leqslant3.8\,\sigma$. However, a full atmospheric retrieval would be needed to provide confidence intervals on the retrieved abundance for water. On individual nights, we find that the predominant signal from water originates from night 4, in effect this means that we were able to detect the signature for water in only $\sim5$\,h of observation in a phase coverage of $\phi = 0.42-0.48$. We find no strong evidence for any phase variability in the water signal over the phase coverage of our observations in our analysis when we split the data into pre- and post-dayside observations. We retrieve an atmospheric scaling factor of $\log(a) = 0.39$ which suggests the model is underestimating the depth of the water lines by $2.5\,\times$, however, this value could be dominated by the mixing of day and night-side emission from the atmosphere due to the $\sim41^{\circ}$ inclination of the planet. This analysis is in agreement with the \textit{L}-band detection from \cite{Lockwood2014} but is strongly in disagreement with the more recent analysis from \cite{Pelletier2021} which finds no evidence for water in the atmosphere of \tauboo b. 

Retrieving an accurate abundance is crucial if we are to understand the C/O ratio \citep[e.g.][]{Madhusudhan2012}, metallicity \citep[e.g.][]{Moses2013} and the physical structure \citep[e.g.][]{Seager1998} of hot Jupiter atmospheres. It is likely that further analysis of the atmosphere of \tauboo b is needed in order to resolve some of the discrepancies that remain over the detection of water. It is also likely that a full atmospheric retrieval is necessary in order to delve deeper into the water features that we have detected from the emergent spectra from \tauboo b in the \textit{Y}-, \textit{J}- and \textit{H}-bands.   

\section*{Acknowledgements}

We would like to acknowledge the use of these python packages that have been used throughout this analysis: NumPy \citep{Numpy}, SciPy \citep{Scipy}, Matplotlib \citep{Matplotlib}, Astropy \citep{Astropy2013,Astropy2018}, emcee \citep{emcee} and corner \citep{corner}. 

We would like to thank the anonymous referee for their comments that have helped to improve this manuscript. MB and SG acknowledge support from the STFC research grant ST/S000631/1. MB acknowledge support from the STFC research grant ST/T000406/1. JLB acknowledges funding from the European
Research Council (ERC) under the European Union's Horizon
2020 research and innovation program under grant agreement No
805445. YZ and IS acknowledge funding from the European Research Council (ERC) under the European Union's Horizon 2020 research and innovation program under grant agreement No. 694513.

CARMENES is an instrument at the Centro Astron\'{o}mico Hispano-Alem\'{a}n (CAHA) at Calar Alto (Almer\'{i}a, Spain), operated jointly by the Junta de Andaluc\'{i}a and the Instituto de Astrof\'{i}sica de Andaluc\'{i}a (CSIC). CARMENES was funded by the Max-Planck-Gesellschaft (MPG), the Consejo Superior de Investigaciones Cient\'{i}ficas (CSIC), the Ministerio de Econom\'{i}a y Competitividad (MINECO) and the European Regional Development Fund (ERDF) through projects FICTS-2011-02, ICTS-2017-07-CAHA-4, and CAHA16-CE-3978, and the members of the CARMENES Consortium (Max-Planck-Institut f\"{u}r Astronomie, Instituto de Astrof\'{i}sica de Andaluc\'{i}a, Landessternwarte K\"{o}nigstuhl, Institut de Ci\`{e}ncies de l'Espai, Institut f\"{u}r Astrophysik G\"{o}ttingen, Universidad Complutense de Madrid, Th\"{u}ringer Landessternwarte Tautenburg, Instituto de Astrof\'{i}sica de Canarias, Hamburger Sternwarte, Centro de Astrobiolog\'{i}a and Centro Astron\'{o}mico Hispano-Alem\'{a}n), with additional contributions by the MINECO, the Deutsche Forschungsgemeinschaft through the Major Research Instrumentation Programme and Research Unit FOR2544 ``Blue Planets around Red Stars", the Klaus Tschira Stiftung, the states of Baden-W\"{u}rttemberg and Niedersachsen, and by the Junta de Andaluc\'{i}a. This work was based on data from the CARMENES data archive at CAB (CSIC-INTA).

\section{Data availability}

The data used in this article is available at http://caha.sdc.cab.inta-csic.es/calto/ with project ID's: 271375, 272952, 294943, 295318, 296169. 




\bibliographystyle{mnras}
\bibliography{papers}



\appendix

\section{MCMC analysis on the individual nights} \label{sec: appendix mcmc ind nights}

\begin{figure}
    \centering
    \includegraphics[width=\columnwidth]{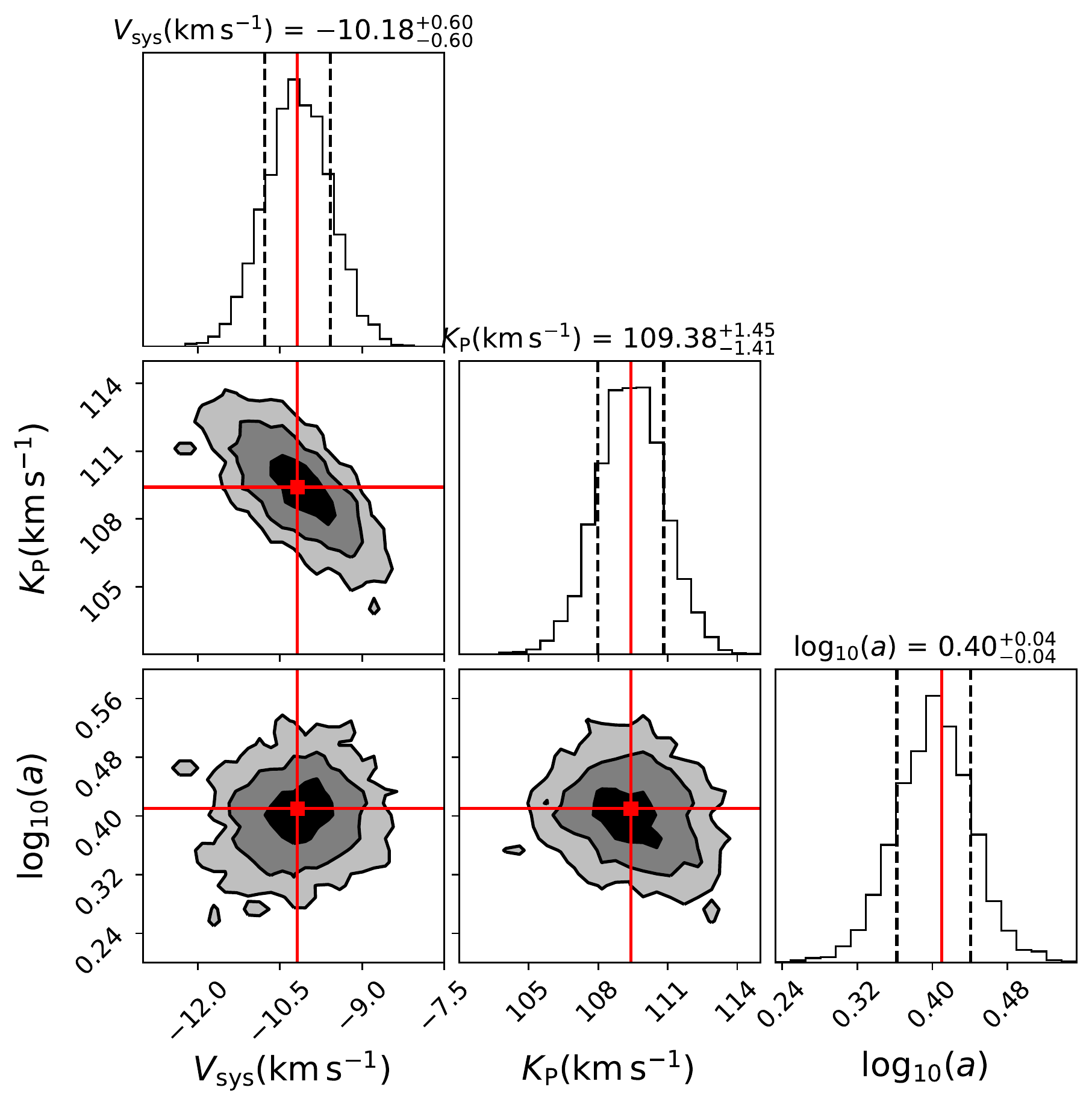}
    \caption{Same as Fig.~\ref{fig: 4 nights corner} and \ref{fig: pre&post corner} but with only spectra from night 4 with the best-fitting combined water model. There is a clear signal from \tauboo b with the retrieved parameters stated above each histogram.}
    \label{fig: combined corner}
\end{figure}

It is clear from Fig.~\ref{fig: pre&post corner} that each nights set of spectra do not contribute equally to the overall water signal from \tauboo b in Fig.~\ref{fig: combined corner}. Here, we repeat the analysis from section~\ref{subsec: mcmc} but instead of combining the nights together, we run an MCMC on the individual nights to assess the contribution from each night. As the pre-superior conjunction data only included night 4, the retrieved parameters will be the same as shown in Fig.~\ref{fig: pre&post corner}.

\begin{table}
\centering
\caption{The median values retrieved from an MCMC analysis for the orbital solution and an atmospheric scaling parameter for each night in the analysis.}
\begin{tabular}{lccc}
\hline
\multicolumn{1}{|l|}{\textbf{Night}}  & \multicolumn{3}{c}{\begin{tabular}[c]{@{}c@{}}\textbf{Retrieved parameters}\\ \textbf{(median)}\end{tabular}} \\ \hline 
\multicolumn{1}{c}{} &$V_{\mathrm{sys}}$ ($\mathrm{km\,s^{-1}})$  &$K_{\mathrm{P}}$ ($\mathrm{km\,s^{-1}}$) & $\log(a)$  \\ \hline 
1 & \multicolumn{1}{c|}{$-15.57^{+21.82}_{-2.70}$} & \multicolumn{1}{c|}{$50.69^{+1.19}_{-0.55}$} & \multicolumn{1}{c|}{$1.71^{+0.06}_{-0.08}$} \\ 
2 & \multicolumn{1}{c|}{$21.44\pm0.19$} & \multicolumn{1}{c|}{$50.03^{+0.05}_{-0.02}$} & \multicolumn{1}{c|}{$1.80\pm0.01$} \\
3 & \multicolumn{1}{c|}{$-10.56^{+10.86}_{-12.85}$} & \multicolumn{1}{c|}{$102.25^{+37.92}_{-43.29}$} & \multicolumn{1}{c|}{$0.27^{+0.26}_{-0.46}$} \\ 
4 & \multicolumn{1}{c|}{$-12.41^{+2.23}_{-2.60}$} & \multicolumn{1}{c|}{$108.22^{+6.36}_{-7.81}$} & \multicolumn{1}{c|}{$0.43^{+0.09}_{-0.15}$} \\ 
5 & \multicolumn{1}{c|}{$-15.56^{+5.67}_{-13.73}$} & \multicolumn{1}{c|}{$110.78^{+8.00}_{-13.95}$} & \multicolumn{1}{c|}{$1.36^{+0.60}_{-1.68}$} \\ \hline
\end{tabular} \label{table: nights mcmc}
\end{table}

In table~\ref{table: nights mcmc} we show the median values retrieved from an MCMC analysis on each night. As described in section~\ref{subsec: mcmc}, the MCMC is set-up with 12 individual walkers each with a chain length of 1000. All of the nights failed to converge to a single solution, however, night 4 performed the best with the main peak in the posterior distribution at the expected radial velocity of the planet. Nights 1 and 2 perform worse than the other nights as those chains settled onto the lower limit of the $K_{\mathrm{P}}$ prior. This is likely an indication that those spectra suffer from some residual tellurics despite the lack of visible highly deviant wavelength columns. Night 5 shows a potentially weak signal for water at the orbital solution of \tauboo b, however, the MCMC did not converge to a single solution with a particularly wide posterior on the scaling factor $\log_{10}(a)$. 

The lack of a water detection in every night except for night 4 is not one of surprise given the variable observing conditions between each night at the Calar Alto site. Given the variable SNR for nights 1, 2, 3 and 5 it is highly likely that these suffered from cloudy conditions intermittently throughout these nights. The variable SNR for nights 1 and 2 combined with the high humidity, which consistently exceeded >85 per cent during the nights, is likely the reason why residual tellurics overpower the weak water signal in this MCMC analysis. Although we can only detect the water signal from the night 4 spectra, it is nevertheless the case that with the addition of nights together helps the water detection by tightening the confidence intervals on the retrieved parameters which is seen in Fig.~\ref{fig: pre&post corner} and \ref{fig: combined corner}. 

\section{Cross-correlation to log likelihood analysis including all nights} \label{app: CC to logl all nights}

\begin{figure}
    \centering
    \includegraphics[width=\columnwidth]{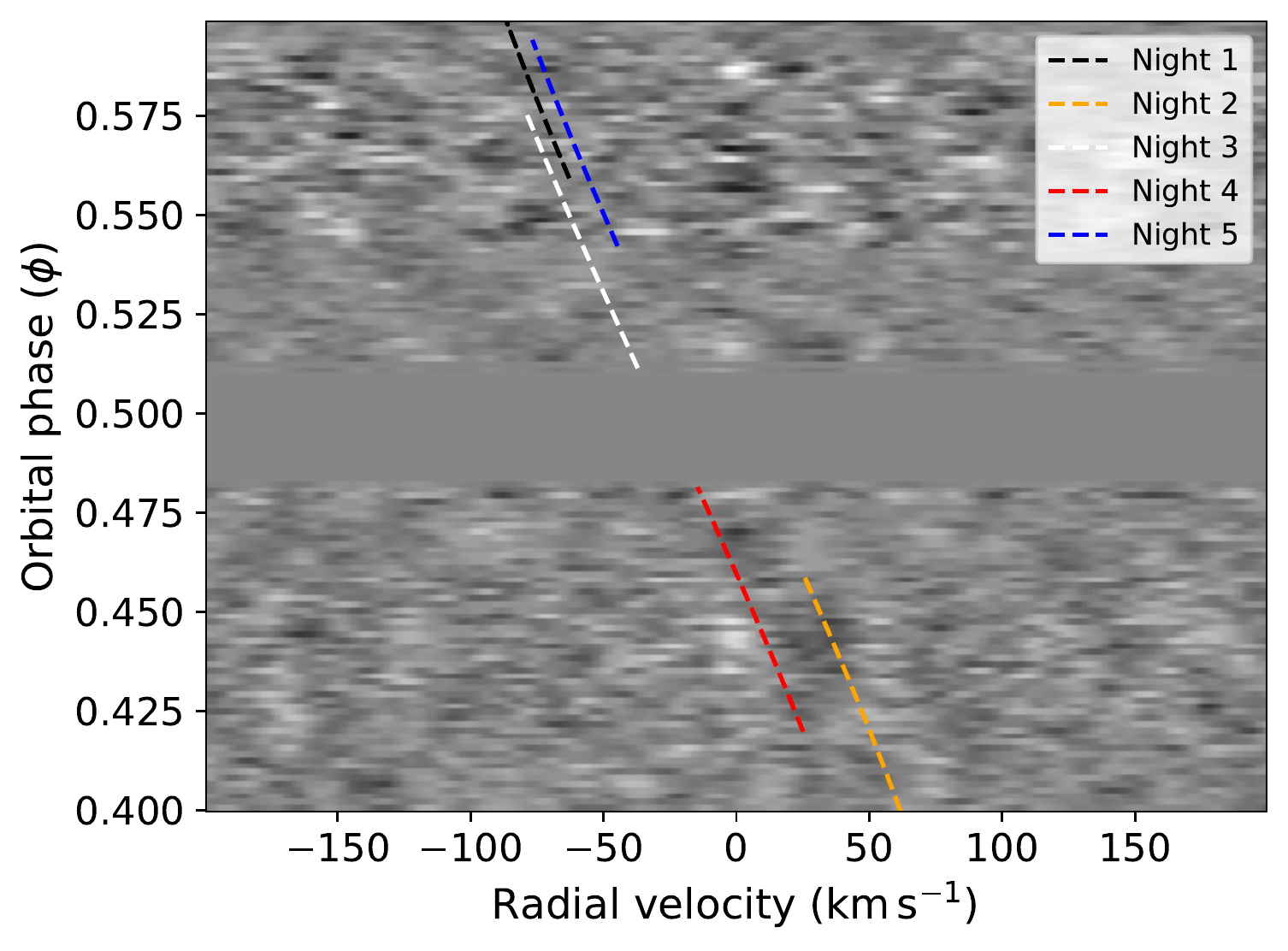}
    \caption{Same as Fig.~\ref{fig: n5 logl ph bin} but with all nights included binned in phase with a resolution of $\Delta\phi=0.0015$.  The black, orange, white, red and blue dashed lines show the expected radial velocity of \tauboo b for nights 1-5, respectively.}
    \label{fig: all nights logl ph bin}
\end{figure}

\begin{figure}
    \centering
    \includegraphics[width=\columnwidth]{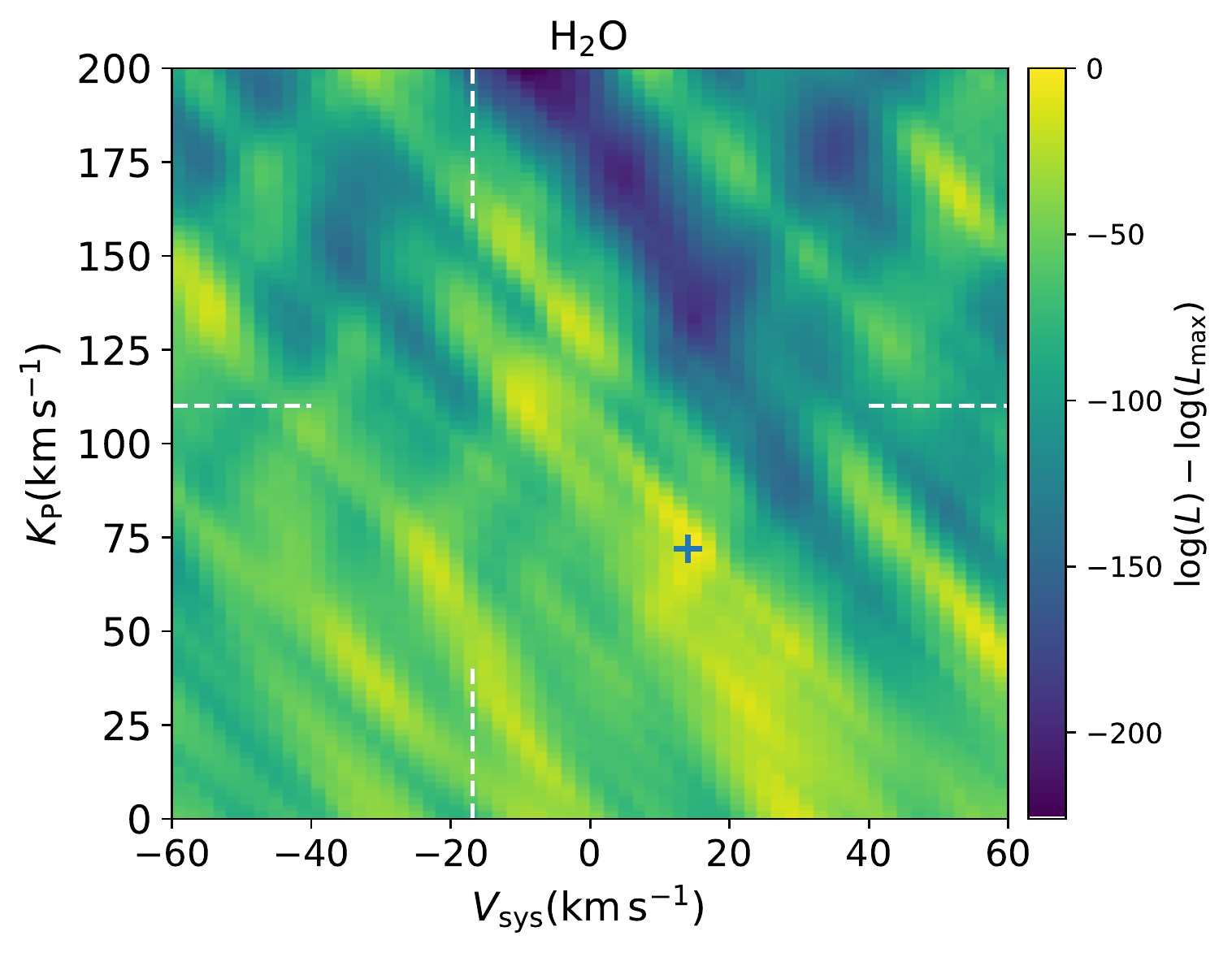}
    \caption{Same as Fig.~\ref{fig: best fit signal from 4 nights} but with the inclusion of night 2. The white dashed lines show the expected position of \tauboo b. The blue cross shows the position of the maximum $\log(L)$ which no longer appears at the position of the detection in Fig.~\ref{fig: best fit signal from 4 nights}.}
    \label{fig: h2o logl all nights }
\end{figure}

\begin{figure}
    \centering
    \includegraphics[width=\columnwidth]{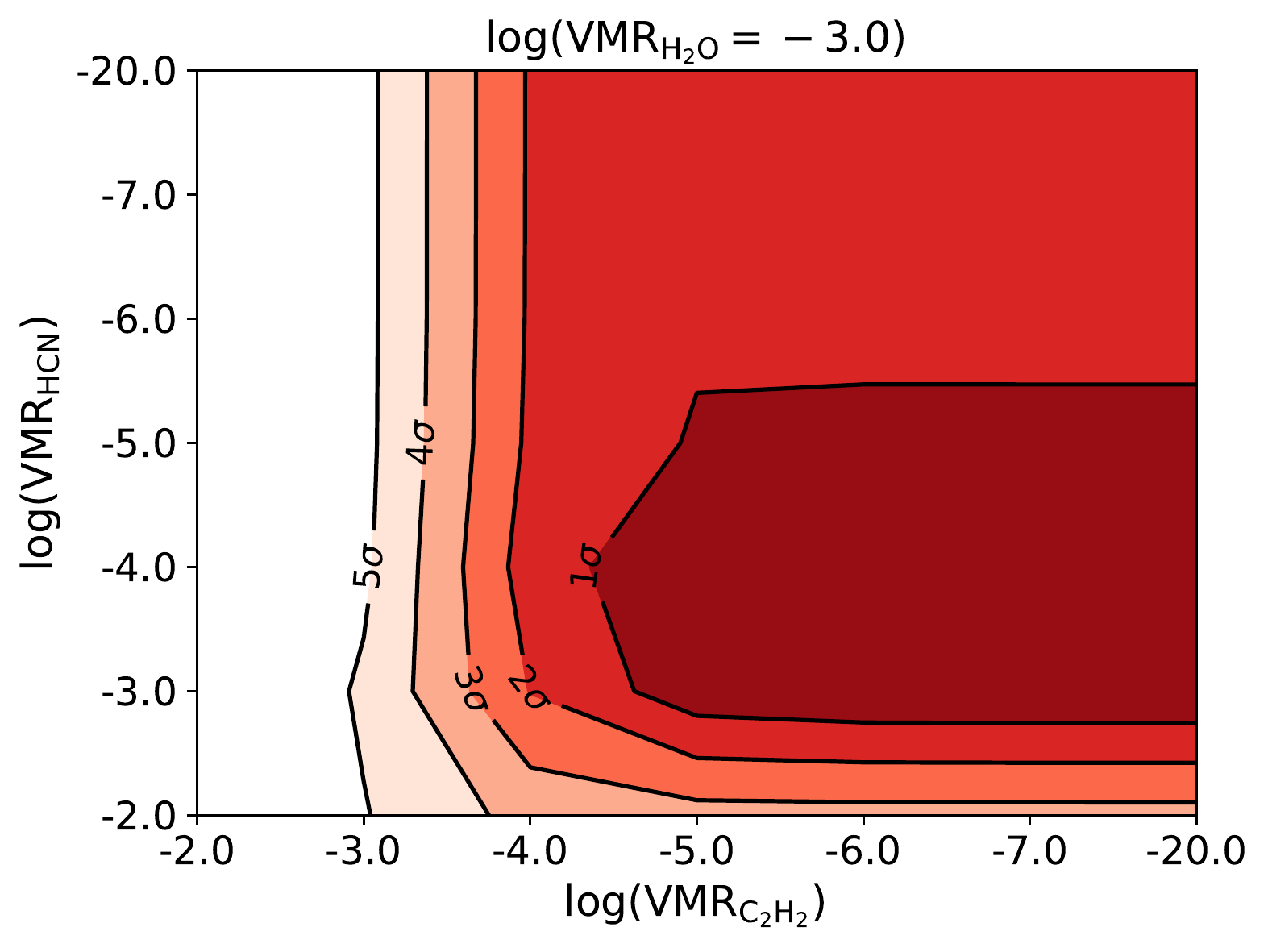}
    \caption{Same as Fig.~\ref{fig: abundance constraints 4 nights} but with the addition of night 2. With the inclusion of these spectra, we see a marginal preference for the addition of HCN at a VMR$=10^{-4}$ in the best-fitting atmospheric model.}
    \label{fig: abundance constraints all nights}
\end{figure}

As shown in the main data analysis (see Fig.~\ref{fig: n5 logl ph bin}), night 2 suffers from strong telluric residuals despite the telluric removal steps and the additional masking of highly deviant spectral channels. However, here, we show how the inclusion of the observations from night 2 (261 additional spectra) affect the results of the analysis from the four other nights. 

Fig.~\ref{fig: all nights logl ph bin} shows all spectra correlated with a pure water spectrum as a function of phase and radial velocity shift from the telluric rest frame. The coloured dashed lines indicate the expected radial velocity trails of \tauboo b during those observations. In Fig.~\ref{fig: h2o logl all nights }, we show the $K_{\mathrm{P}}-V_{\mathrm{sys}}$ velocity map with the inclusion of night 2. It is clear that the telluric noise from these spectra overwhelms the planet's water signal as seen in Fig.~\ref{fig: best fit signal from 4 nights} with the retrieved $\log(L_{\mathrm{max}})$ (shown as the blue cross) shifting beyond the expected orbital velocity of the planet. However, as shown in Fig.~\ref{fig: combined corner}, the MCMC still converges onto a local maximum at the expected orbital solution of the planet. Although this is a case of the MCMC algorithm converging onto a local maximum before exploring the wider parameter space, it is nevertheless showing that there is still a detectable signal of water from the atmosphere in the local vicinity of expected orbital solution. If we use the retrieved parameters from the MCMC analysis, we find a marginal preference of $1.1\,\sigma$ for the inclusion of HCN in the atmospheric models at an abundance of VMR$=10^{-4}$. In Fig.~\ref{fig: abundance constraints all nights}, we show the updated abundance constraints on HCN and C$_{2}$H$_{2}$ for the inclusion of night 2 spectra. With these results we can lower the $3\,\sigma$ upper limits on the these species at $\log(\mathrm{VMR}) \approx -2.0\,\mathrm{and}\,\approx -4.0$ for HCN and C$_{2}$H$_{2}$, respectively.   


\bsp	
\label{lastpage}
\end{document}